%% file: paper.tex
\documentclass[final,5p,times,twocolumn]{elsarticle}
\usepackage{url}

\usepackage{amssymb}

\journal{Control Engineering Practice}

\begin{document}

\begin{frontmatter}




\title{Flight control of tethered kites in autonomous pumping cycles for
airborne wind energy}



\author{Michael Erhard\corref{cor1}}
\ead{michael.erhard@skysails.de}
\cortext[cor1]{Corresponding author}

\author{Hans Strauch\corref{}}

\address{SkySails GmbH, Luisenweg 40, D-20537 Hamburg, Germany}

\begin{abstract}
Energy harvesting based on tethered kites benefits from
exploiting higher wind speeds at higher altitudes.
The setup considered in this paper is based on a pumping cycle.
It generates energy by winching out at high tether forces, driving an electrical
generator while flying crosswind. Then it winches in at a stationary neutral
position, thus leaving a net amount of generated energy.

The focus of this paper is put on the flight control design, which implements 
an accurate direction control towards target points and allows for a flight with an
eight-down pattern. An extended overview on the control system
approach, as well as details of each element of the flight controller, are
presented.
The control architecture is motivated by a simple, yet comprehensive model
for the kite dynamics.  

In addition, winch strategies based on an optimization scheme are presented.
In order to demonstrate the real world functionality of the presented
algorithms, flight data from a fully automated pumping-cycle operation of a
small-scale prototype are given. The setup is based on a 30\,m$^2$ kite linked to
a ground-based 50\,kW electrical motor/generator by a single line.
\end{abstract}


\end{frontmatter}


\input{introduction.tex}
\input{setup.tex}
\input{power_pattern.tex}
\input{plant.tex}
\input{control_overview.tex}
\input{control_psidot.tex}
\input{control_psi.tex}
\input{control_targetpoint.tex}
\input{control_cycle.tex}
\input{control_winch.tex}
\input{results.tex}
\input{summary.tex}
%

\appendix
\input{appendix.tex}





\input{paper.bbl}
\end{document}

%% file: introduction.tex
\section{Introduction} 
More than thirty years ago \cite{Loyd1980} energy generation using tethered wings
has been proposed for the first time. Since then a great interest in this kind of
renewable energy source has emerged, especially during the last decade. 
The application of tethered wings or kites appear very attractive, as they
combine high achievable forces in crosswind flight together with the possibility of
easily venturing into higher flight altitudes thereby taking advantage of the higher wind speeds.

The different concepts can be grouped together by using the term 'airborne
wind energy', for an overview see e.g.~\cite{Fagiano2012a}.
An extended summary on geometries, theory oriented research activities, realized
prototype systems and planned setups can be found in the recent textbook on
airborne wind energy \cite{Ahrens2013}.

The economic operation of airborne wind energy plants demands for reliable and
fully automatic operation of the power generation process. Thus, numerous
theoretical control proposals \cite{Diehl2001a, Ilzhoefer2007,
Fagiano2011, Baayen2012, Lellis2013a} as well as experimental implementations
have been published \cite{Erhard2012a, Jehle2012, Jehle2014, Fagiano2013a}.
However, the robust autonomous operation of
complete energy production cycles turns out to be quite challenging,
especially as optimization of energy output, i.e.~performance and robustness,
often appear as opposing design prerequisites. Hence, a design process for the control system,
which takes into account real world circumstances to the necessary degree, is required. We are convinced,
that simplicity, separation of problems and modular structure, grounded in a clear understanding of the
physical basis of the controlled plant, are keys to success in mastering the high
perturbations and significant uncertainties, which are inevitably coming
along with the natural energy resource wind. 

This paper will report on the control system for
complete autonomous power cycles with a small-scale 50\,kW prototype system
using a 30\,m$^2$ kite. 
The focus is put on flight control of efficient dynamical pattern-eight
trajectories, which are crucial in order to obtain an optimal power
generation output.
A distinguishing feature of the pattern-eight flight trajectories is the option
of flying them in two ways.
From the practical point of view, one would prefer the so called eight-down
trajectories as those significantly decrease force variability.
This allows
for a broader operational range of wind conditions and thus
increases the average power output.
However, the performance advantage comes along with drawbacks of
temporarily flying ahead towards the surface and with the need for proper curve flights,
which pose special requirements to the flight control system. As a consequence,
the previously published control system \cite{Erhard2012a} was extended in
order to combine it with target point concepts similar to \cite{Fagiano2013a} and
\cite{Vlugt2013a}. 
For the overall power generation control, a compact description with three
states and simple winch control strategies for the different phases have been
added, which already yield remarkable results.

In this manuscript, the complete control design shall be presented, based on
the equations of motion of a model \cite{Erhard2012a}, which
describes the steering behavior of the kite as well as the kinematics, and
has been extended for changing tether lengths in \cite{Erhard2013b}.
The single design steps towards a robust pattern eight-down flight are discussed
in detail and the applicability is illustrated by the discussion of real flight
data results.

The paper is organized as follows: starting with a brief summary of the
system setup and power generation principle in Sect.~\ref{sec:setup}, the pattern
eight-down flight and control prerequisites are motivated in
Sect.~\ref{sec:control_task}.
After summarizing the equations of motion in Sect.~\ref{sec:plant}, an overview on the
complete control setup is presented in Sect.~\ref{sec:control_overview}.
Sections \ref{sec:control_psidot}--\ref{sec:control_winch} present details of
the single controller parts and illustrate their principle of operation by
discussing experimental flight data in Sect.~\ref{sec:results}.
A summary and outlook is given in Sect.~\ref{sec:summary}.

%% file: setup.tex
\section{Implemented prototype and power generation}
\label{sec:setup}
In this section, a general overview on the architecture and the operation principle for
power generation will be given. An extended description of involved
components and background information can be found in \cite{Fritz2013a}.
\subsection{Setup}
\begin{figure}
  \centering
  \includegraphics[width=8.8cm]{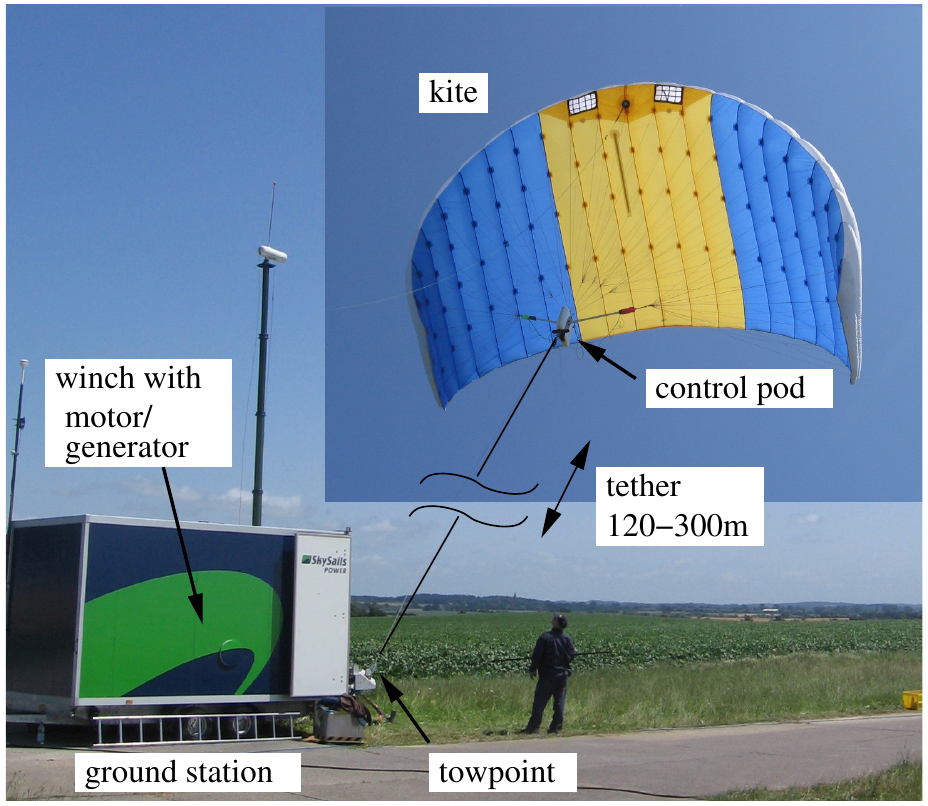}
  \caption{Small scale prototype system for kites
  of sizes ranging from 20--40\,m$^2$ (30\,m$^2$ shown here). 
  The main winch with motor/generator is located in the ground station. A
  tether line of length typically in the range 150--300\,m transfers the forces
  from the flying system. A distinguishing feature of the latter is the control
  pod located under the kite, which allows for a single towing rope. The
  actuator in the control pod pulls certain lines in order to steer the kite.}
  \label{fig:setup}
\end{figure}

A picture of the small-scale prototype is shown in
Fig.~\ref{fig:setup}. 
The ram-air kite of 30\,m$^2$ is controlled by steering
lines, which are pulled by an actuator placed in a control pod. The pod is directly
located under the kite. 
This geometry allows for a {\em single} main towing line, consisting of 6\,mm
diameter high-performance Dyneema\textsuperscript{\textregistered}
rope, which connects the flying system to the ground station and transfers the aerodynamic forces. 
The prototype features 300\,m of tether length on the main
winch, which is attached to a 50\,kW electrical motor/generator-combination.

In order to support research and development projects, the prototype is
equipped with several sensors. Although the specific choice of sensors and
signal preprocessing is important for the whole control design, a detailed discussion
would go beyond the scope of this paper with its emphasis on control.
However in order to allow for a proper understanding of the subsequent
sections, a short summary on the most important sensors is given in Table
\ref{tab:sensors}.
\begin{table}
  \centering
  \begin{tabular}{p{1cm}p{6.8cm}}
  \hline
  \multicolumn{2}{c}{\bf Control pod sensors}\\
  \hline\hline
  $\psi_{\rm m}$ & Orientation angle w.r.t.~wind, determined by a 6-DOF (3 turn
  rate sensors and 3 accelerometers) inertial measurement unit (IMU) with
  semiconductor MEMS sensors. The fusion algorithm, based on complementary
  filtering, also takes into account the wind direction (see below $v_{\rm w}$).\\
  $\dot{\psi}_{\rm m}^\prime$ & Turn rate around yaw axis measured by the
  corresponding turn rate sensor of the IMU.\\
  $v_{\rm a}$ & Air path speed of the flying system measured by a propeller
  anemometer located at the control pod.\\
  $F_{\rm pod}$ & Tether force measured by a strain gauge.\\
  \hline
  \multicolumn{2}{c}{\bf Ground unit sensors}\\
  \hline\hline
  $\varphi_{\rm m},\vartheta_{\rm m}$ & Mechanically sensed direction of
  the tether, referenced to the wind direction (see $v_{\rm w}$).\\
  $l$ & Tether length based on rotations of multi-turn encoder attached to
  the drum.\\
  $v_{\rm w}$ & Wind speed measured by a 2d ultrasonic anemometer of the
  ground station, mounted at 5\,m altitude. The wind direction is used as reference for the towpoint readings
  and the inertial measurement unit of the control pod. 
  It is worth mentioning, that an estimation algorithm for the mean wind speed
  and wind direction at flight altitude may depend on weather conditions. The use of an estimator can
  significantly improve the robustness and it should be used instead of the
  anemometer reading. However, even in the latter case the anemometer is still
  important as initial condition and validation input to the estimator.\\
  \hline
  \end{tabular}
  \caption{Overview on sensors and origin of measured values, restricted to
  quantities discussed in this paper. Proper definitions of the values can be
  found in the respective sections of the controller description.}  
  \label{tab:sensors}
\end{table}
For a detailed overview on the sensors for flight control of
tethered kites, the interested reader is referred to \cite{Erhard2013a} and to
\cite{Fagiano2014a}, \cite{Ranneberg2013a} for application examples of fusion
algorithms. In the following, measured sensor quantities are indicated by
the subscript '$_{\rm m}$'.
\subsection{Power generation cycle}
\begin{figure}
  \centering
  \includegraphics[width=8.8cm]{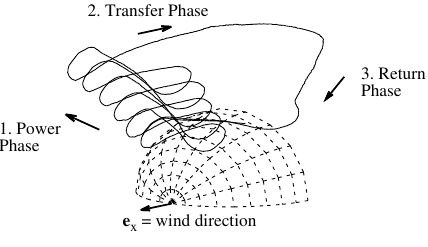}
  \caption{Flight trajectory for the power generation cycle. 3d view of
  experimental flight data.}
  \label{fig:flight_path}
\end{figure}

This subsection focuses on the applied principles of  power generation. A typical 
flight trajectory during operation is sketched in Fig.~\ref{fig:flight_path}.
The power generation is done in cycles, which consist of the
following three phases:
\begin{enumerate}
  \item In the {\em power generation phase}, the kite is flown dynamically in
  pattern-eight configuration, which induces high line forces. Meanwhile the
  line is winched out, driving an electrical generator producing energy.
  \item When a certain line length is reached, the {\em transfer phase} brings
  the kite to a neutral position. The heading is against the
  wind resulting in a low line force.
  \item During the {\em return phase}, the line is winched in, operating the
  generator as motor while the kite is kept at a neutral wind window position.
  This phase consumes a certain amount of the energy produced in phase 1. As the tether
  force at neutral position is much lower than during dynamic flight, only a
  minor fraction of the generated energy of phase 1 is needed leaving a
  considerable net amount of generated energy.
  When the lower line length threshold is reached, the whole cycle repeats
  starting at (1). 
\end{enumerate}
This periodic winching process is also called pumping cycle
or yo-yo operation configuration.

Finally, it should be remarked that the kite is flown
with constant angle of attack during all phases and there is no de-powering feature for the
return phase implemented as e.g.~in \cite{Vlugt2013a}. Therefore, the return
phase is accomplished by winching the kite {\rm directly against the wind}. 
At first sight, this
strategy seems to be inefficient as it suggests slow winching
speed in order to keep down tether forces.
However, rather contrary to intuition, the air flow at the
kite, and subsequently the tether forces, are even reduced by increasing the winch
speed as shown in Sect.~\ref{sec:plant_winching}, making this power
generation scheme competitive. 
The extension of the scheme by variation of the angle of attack would
demand for an additional control actuator, which increases complexity and weight
of the airborne system. Evaluation of the performance gains versus costs is
subject to current theoretical and experimental research activities.

%% file: power_pattern.tex
\section{Effective power pattern}
\label{sec:control_task}
\begin{figure}
  \centering
  \includegraphics[width=8.8cm]{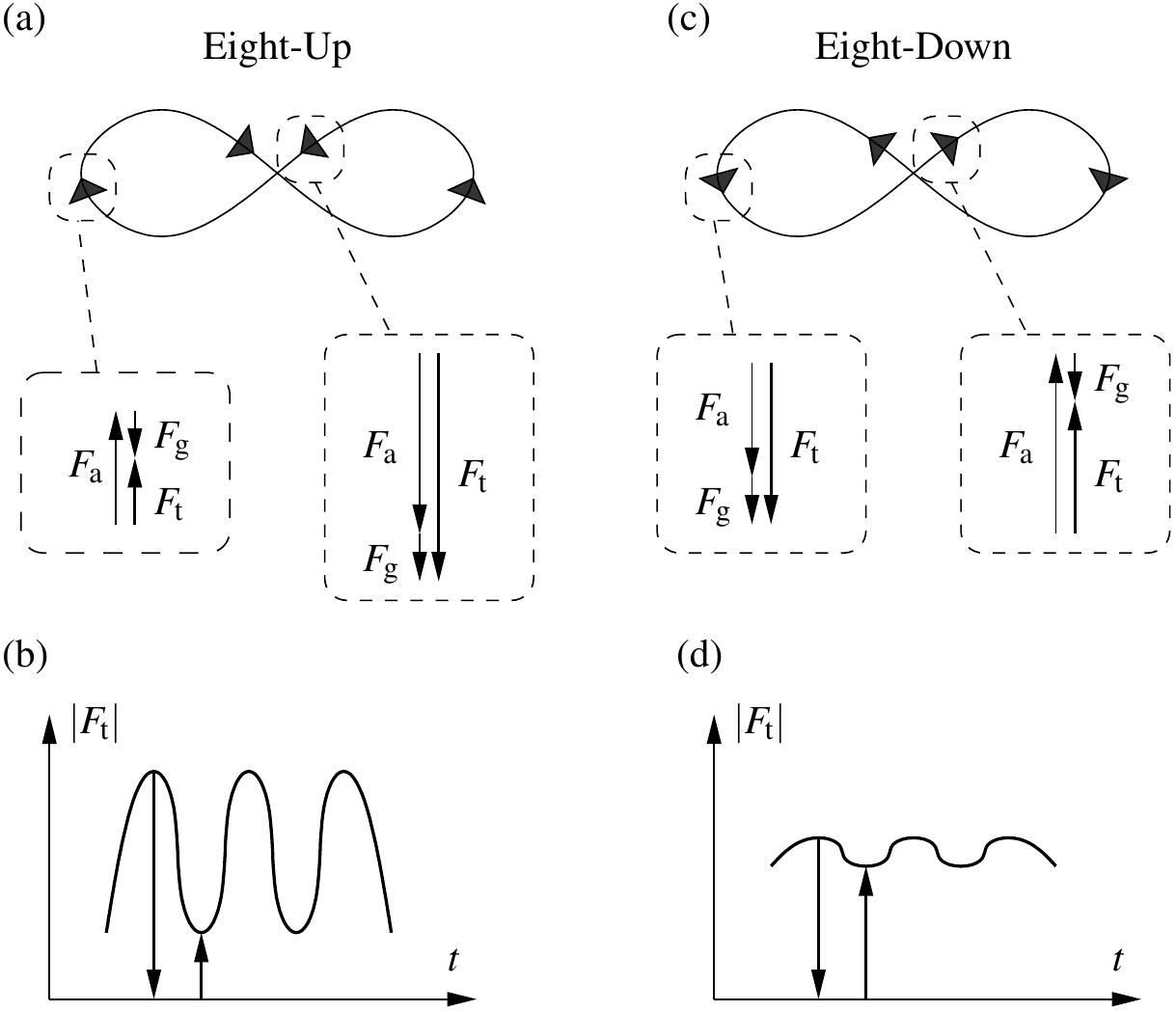}
  \caption{Principle of pattern-eight flight. Note that the eight can be flown
  in two ways indicated by the triangular shaped arrows on the eight
  trajectories.
  The eight-up pattern (a) suffers from low aerodynamic forces $F_{\rm a}$ in the outer
  regions further reduced by gravity $F_{\rm g}$, leading to low total forces
  $F_{\rm t}$. In the center region, the higher aerodynamic forces are even
  increased by gravity leading to huge variations of the force with respect to
  time as shown in (b). For the eight-down variant (c), the variations of
  gravity and aerodynamic forces compensate partly, leading to a more regular
  total force, compare (d).}
  \label{fig:pattern_eight}
\end{figure}

Effective power generation with tethered kites make use of the 
huge traction forces, which are generated by dynamical pattern-eight flight.
An important distinguishing feature is, that the pattern-eight can be flown in
two ways as illustrated in Fig.~\ref{fig:pattern_eight}.
Note the triangles drawn on the trajectories in the figures indicating the
flight directions. Comparing the eight-up and eight-down configurations
with respect to maximum energy generation, the eight-down variant is clearly
favorable due to the better compensation of gravity by aerodynamic forces as
shall be explained in the following.
The highest aerodynamic forces occur in the center of the wind window.
The eight-up pattern, shown on the left hand side, significantly suffers from
flying up against gravity in the outer regions with low aerodynamic forces.
In order to improve performance, one can take advantage of gravity in the outer
regions with lower aerodynamic forces by flying down and using the high-force
region in the center to fly up against gravity. This is achieved by the
eight-down pattern shown on the right hand side.
A further advantage of these eights-down is the significantly reduced variation
in tether force as well as a reduction of the required minimum wind speed for
stable operation. Both benefits make the pattern-eight-down concept very
attractive for power generation, especially for pumping cycle concepts,
based on power generation by winching, which leads to a significant reduction of the apparent
wind speed.

Note, that in contrast to eight-ups, which could be controlled by a quite simple
approach \cite{Erhard2012a,Erhard2013b}, the given challenge of automated
pattern-eight-down flight demands for a certain level of trajectory
guidance. We consider the following two properties as essential.
First, the flight towards a target point \cite{Fagiano2013a} must be controlled
quite accurately in order to always keep the kite in the desired region of the
wind window and especially in order to compensate for perturbations due to side
gusts.
Second, the curve flight, which is basically steered by a trapezoidal signal on
the actuator, must be well engineered because a major part of the steering speed of
the control actuator has to be used explicitly to allow for reasonable small
curve radii. In addition, the final target direction of a curve flight should
be reached quite accurately with absence of significant overshoots.

Although gravity effects are important for the
specific choice of the flight pattern for performance reasons, they are
neglected in the following as they could be regarded as uncertainties in the
dynamics, which are compensated by the control feedbacks. A thorough modeling
in order to cover low wind performance is subject to current research activities
and will be published elsewhere.

%% file: plant.tex
\section{Plant description}
\label{sec:plant}
\begin{figure}
  \centering
  \includegraphics[width=4cm]{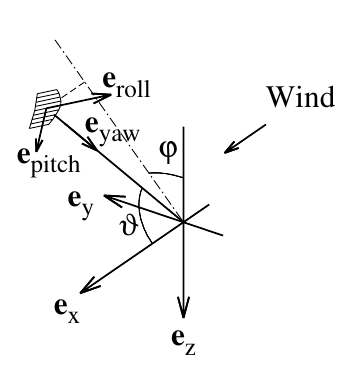}
  \caption{Coordinate system. The distance between origin and kite is given by
  the tether length $l$.}
  \label{fig:coordinates}
\end{figure}

In this section, the dynamic model of the system shall be presented. The
extended derivation of the model as well as system identification and
comparing the model to experimental data is done in \cite{Erhard2013b}.
The model is based on four state variables 
${\bf x}=\left[\varphi,\vartheta,\psi,l\right]$ as shown in
Fig.~\ref{fig:coordinates}.
The position of the kite $\bf{r}$ is given in polar coordinates by the angles
$\varphi,\vartheta$ in combination with the tether length $l$ and could be
written as
\begin{equation}
  {\bf r} = l\, 
  \left(\begin{array}{c} 
    \cos\vartheta \\ 
    \sin\varphi \sin\vartheta \\ 
    - \cos\varphi \sin\vartheta
  \end{array}
  \right)
\end{equation}
The orientation of the kite is parameterized by the angle $\psi$, which could be
defined as angle to the wind, i.e.~$\psi=0$ corresponds to a heading directly
against the wind.
The tether of the model could be regarded as a rigid rod, hence the complete
orientation of the kite is determined by the angles $\varphi,\vartheta,\psi$.
An ambient constant and homogeneous wind field with wind speed $v_{\rm
w}$ along the $x$-direction is assumed.
\subsection{Model assumptions}
In order to derive the equations of motion, some assumptions are made, which
shall be summarized and justified in the following:
\begin{itemize}
  \item The aerodynamic forces are assumed to be large compared to
  system masses. This allows for some simplifications. First, the rope can be
  implemented as simple tether (mass-less and infinitely thin rod). Second, all
  acceleration effects can be neglected assuming the system is always in an
  equilibrium of forces. Especially the latter assumption significantly reduces
  the equations of motion from second to a first order system.
  \item The aerodynamics of the kite is reduced to the glide ratio number $E$,
  which describes the ratio of the air flow between roll- and yaw axis. Further,
  it is assumed, that there is no air flow component in pitch-direction (no
  side-slip).
  \item The steering behavior of the kite is described by a simple turn-rate
  law, which has been introduced empirically and shown
  experimentally \cite{Erhard2012a}, \cite{Jehle2014}, but also can be nicely
  derived from first principles \cite{Fagiano2013a}.
\end{itemize}
\subsection{Equations of motion}
The kite system input is described by the control input vector ${\bf
u}=\left[\delta,v_{\rm winch}\right]$.
The deflection $\delta$ determines the steering input applied by the actuator
of the control pod to the kite.
As winch dynamic is not subject to this paper, it is assumed, that $v_{\rm
winch}$ directly determines the change of tether length $l$.

Taking into account the assumptions of the previous subsection, the following 
equations of motion can be derived:
\begin{eqnarray}
  \dot{\psi} &=& g_{\rm k}\, v_{\rm a}\, \delta + \dot{\varphi}
  \cos\vartheta \label{eq:eqm_psi}\\
  \dot{\vartheta} &=& \frac{v_{\rm a}}{l}
  \left(\cos\psi - \frac{\tan\vartheta}{E}\right) -
  \frac{\dot{l}}{l} \tan\vartheta\label{eq:eqm_theta}\\
  \dot{\varphi} &=& -\frac{v_{\rm a}}{l \sin\vartheta} \sin\psi
  \label{eq:eqm_phi}\\
  \dot{l} &=& v_{\rm winch}\label{eq:eqm_l}
\end{eqnarray}
The system parameter $g_{\rm k}$ quantifies the response of the kite due to a
steering deflection.
The value $v_{\rm a}$ represents the air path velocity at the kite as it
is measured by an on-board anemometer located in the control pod.
The same assumptions yield the following relation
\begin{equation}
  v_{\rm a} = v_{\rm w} E \cos\vartheta - \dot{l} E.\label{eq:eqm_va}
\end{equation}
In order to get the final result for the complete set of equations of motion,
one would usually insert (\ref{eq:eqm_va}) into
(\ref{eq:eqm_psi})--(\ref{eq:eqm_phi}). This is also done explicitly or
implicitly when using these equations for numerical simulations.
However, there is a major reason, why this is done only partially here.
As we aim at designing a control algorithm for a real system, the capability of reliably
measuring certain quantities becomes important, as opposed to
a pure simulation.
Measuring $v_{\rm a}$ turns out to be comparably easy by an
on-board anemometer in the control pod, while the determination of $v_{\rm w}$
is quite involved. The straight-forward approach of using an
anemometer next to the ground station would require, that the wind field is
constant and homogeneous. But this is definitely not the case for
airborne wind energy devices making use of higher wind speeds at
higher altitudes. 
Thus, knowing $v_{\rm a}$ by measurement instead of using
(\ref{eq:eqm_va}) could be regarded as kind of generalization of the equations
of motion by taking into account local (measured) effects.
Hence, using $v_{\rm a}$ instead of $v_{\rm w}$, whenever possible, is preferred
in order to enhance the robustness of the control system.

For constant tether lengths $\dot{l}=0$, it can be shown that $\vartheta \leq
\arctan E$ and thus (\ref{eq:eqm_psi})--(\ref{eq:eqm_phi}) could be used with
$v_{\rm a}$ as measurement input.
However, as soon as winching in is allowed, $\vartheta=\pi/2$ is in the usual
operation condition and (\ref{eq:eqm_theta}) is no longer defined due to the
singularity in $\tan \vartheta$. 
In order to resolve this issue, (\ref{eq:eqm_va}) is inserted into
(\ref{eq:eqm_theta})--(\ref{eq:eqm_phi}). The resulting complete set of
equations of motion reads then:
\begin{eqnarray}
  \dot{\psi} &=& g_{\rm k}\, v_{\rm a}\, \delta + \dot{\varphi}
  \cos\vartheta \label{eq:eqm_psi2}\\
  \dot{\vartheta} &=& \frac{v_{\rm w}}{l}
  (E\cos\vartheta\cos\psi - \sin\vartheta) -
  \frac{\dot{l}}{l} E \cos\psi\label{eq:eqm_theta2}\\
  \dot{\varphi} &=& -\frac{v_{\rm w} E \cos\vartheta - \dot{l}E}{l
  \sin\vartheta} \sin\psi\label{eq:eqm_phi2}\\
  \dot{l} &=& v_{\rm winch}\label{eq:eqm_l2}
\end{eqnarray}
\subsection{Motion on a sphere and crossterm}
\label{sec:plant_crossterm}
Due to the motion on a sphere, covering significant angular ranges in short time,
the curvature of the state space has certain effects on the kinematics.   
As the inertial turn-rate sensor is aligned along the yaw axes, one would
obtain a turn rate of $\dot{\psi}_{\rm m}^\prime=0$ for the tethered
motion in absence of other external forces.
For a steering deflection, the turn rate is described by the turn-rate law
as $\dot{\psi}_{\rm m}^\prime=g_{\rm k}\, v_{\rm a}\, \delta $.
Comparing this expression to (\ref{eq:eqm_psi})
\begin{equation}
  \dot{\psi} = g_{\rm k}\, v_{\rm a}\, \delta + \dot{\varphi}\cos\vartheta 
  \label{eq:psidot2}
\end{equation}
one recognizes the term $\dot{\varphi}\cos\vartheta$, which 
implements a cross-coupling between $\psi$ and
$\varphi$,$\vartheta$. This term is defined as 
crossterm
\begin{equation}
  \dot{\psi}_{\rm ct} \doteq \dot{\varphi} \cos\vartheta
  \label{eq:psidot_ct}
\end{equation}
Equation (\ref{eq:psidot2}) then reads
\begin{equation}
  \dot{\psi} = \dot{\psi}_{\rm m}^\prime + \dot{\psi}_{\rm ct}
\end{equation}
In other words, the crossterm $\dot{\psi}_{\rm ct}$ represents the difference
between the turn rate w.r.t.~the inertial system $\dot{\psi}_{\rm m}^\prime$ and
the time derivative of $\dot{\psi}=d/dt \psi$.
Note, that the prime is always used to indicate turn rates w.r.t.~an 
inertial system.

In a cascaded controller topology one would like to setup the chain as $\psi_{\rm
s} \rightarrow \dot{\psi}_{\rm s} \rightarrow \delta$ and implement both
controllers as two cascaded SISO blocks for simplicity and robustness reasons. This could be
done by assuming $\dot{\psi}_{\rm s} \approx \dot{\psi}_{\rm s}^\prime$, which
is an appropriate approximation in some other flight applications. 
However, for highly dynamical pattern flight, consideration of the crossterm is
necessary and can be achieved by using (\ref{eq:psidot2}) explicitly in the controller
design. 
For simulations $\dot{\psi}_{\rm ct}=\dot{\varphi}\cos\vartheta$
 could be used to accomplish the task. Unfortunately, measuring $\dot{\varphi}$
accurately and reliably turns out to be cumbersome, therefore a quantity based
on controller states and sensor values, rather than derivatives of sensor measurements, is preferred.
Using the model relation (\ref{eq:eqm_phi}) yields:
\begin{equation}
  \dot{\psi}_{\rm ct} = \dot{\varphi} \cos\vartheta = -\frac{v_{\rm a}}{l
  \tan\vartheta}\sin\psi \label{eq:crossterm_operational}
\end{equation}
Details on controller implementations and performance discussions will be
presented in Sect.~\ref{sec:control_psi}.
\subsection{Flight direction}
\label{sec:flight_direction}
\begin{figure}
  \centering
  \includegraphics[width=6.5cm]{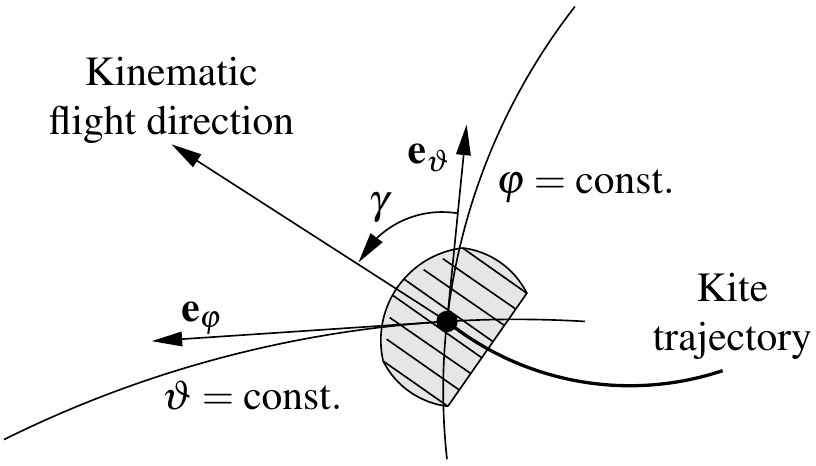}
  \caption{Definition of the flight direction in the
  spherical coordinate system. The angle $\gamma$ is defined with respect
  to the two orthogonal vectors ${\bf e}_{\varphi}$ and ${\bf e}_{\vartheta}$,
  which are locally defined by $\vartheta={\rm const.}$ and $\varphi={\rm
  const.}$, respectively.}
  \label{fig:flight_direction}
\end{figure}

In order to navigate on the sphere, the flight direction $\gamma$ is defined as
follows:
\begin{equation}
  \gamma \doteq \arctan(-\dot{\varphi}
  \sin\vartheta,\dot{\vartheta})\label{eq:def_gamma}
\end{equation}
As depicted in Fig.~\ref{fig:flight_direction}, $\gamma$ denotes the angle
between the 'latitude' line with $\varphi={\rm const}$ through the
current position ${\bf r}$ and the direction of the kinematic motion
$\dot{\bf r}$.
Direct measurement of the flight direction is subject to high noise
caused by the time derivatives of $\varphi_{\rm m}$ and $ \vartheta_{\rm m}$. It is also
sensitive to slack line effects in the towing rope. Therefore directly
processing  $\gamma$ in a control loop, where $\gamma$ is computed from
sensor values using (\ref{eq:def_gamma}), should be
avoided.
An alternative is to control the flight direction indirectly by $\psi$.

In order to obtain the corresponding relation,
(\ref{eq:eqm_theta2}) and (\ref{eq:eqm_phi2}) are inserted into
(\ref{eq:def_gamma})
\begin{equation}
  \gamma = \arctan\left(\sin\psi,
  \cos\psi-\frac{1}{E}\frac{v_{\rm w}\sin\vartheta}{(v_{\rm w}\cos\vartheta - \dot{l})}\right)\label{eq:flight_direction}
\end{equation}
The inversion of this relation is done by using the relation
\begin{equation}
  \psi = \arctan (r \sin\gamma, c_1 + r \cos\gamma) \label{eq:psi_from_gamma}
\end{equation}
were the quantities $r$ and $c_1$ can be determined as:
\begin{eqnarray}
  c_1 &=& \frac{1}{E}\frac{v_{\rm w}\sin\vartheta}{(v_{\rm w}\cos\vartheta -
  \dot{l})}\\
  r &=& \sqrt{1-c_1^2 \sin^2 \gamma} - c_1 \cos\gamma
\end{eqnarray}
The difference between $\gamma$ and $\psi$, which is referenced to the air flow
$\dot{\bf r}-\dot{l}{\bf e}_{\rm yaw}-v_{\rm w} {\bf e}_{\rm x}$, is thus
determined by the background wind vector. In crosswind flight with
$|\dot{\bf r}| \gg v_{\rm w}$, the difference is small compared to the
range of directions in the pattern-eight.
For an accurate direction control, this difference should be considered,
however.
\subsection{Winching}
\label{sec:plant_winching}
\begin{figure}[h]
  \centering
  \includegraphics[width=8.8cm]{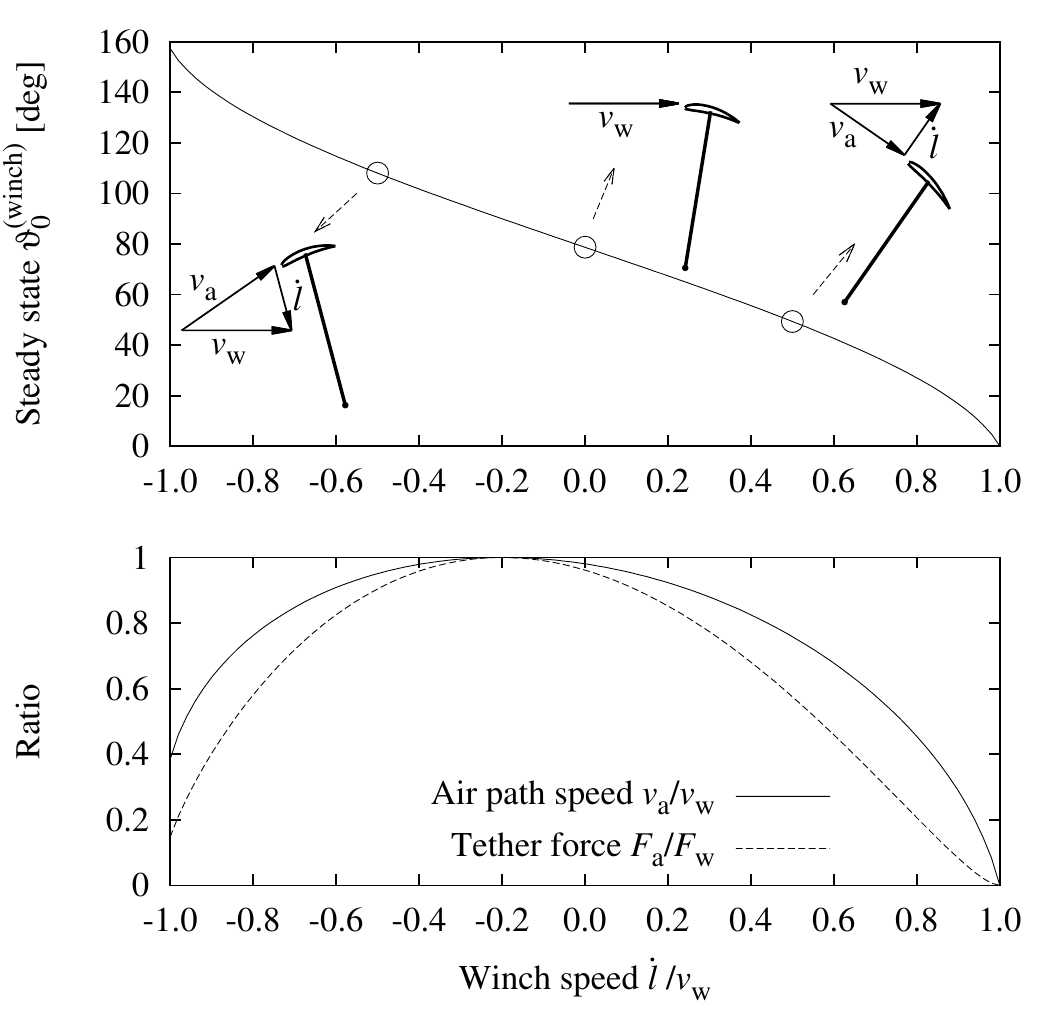}
  \caption{Steady state wind window angle $\vartheta_0^{\rm
  (winch)}$ as function of winch speed $\dot{l}$ for $\psi=0$ and $E=5.0$
  (upper graph).
  Winching in (out) leads to a increasing (decreasing) wind window angle $\vartheta_0^{\rm
  (winch)}$. In the depictive description one obtains a coming forth (falling
  back) of the kite as indicated by the subfigures for $\dot{l}=\pm 0.5 v_{\rm
  w}$.
  The lower figure shows the air path speeds and tether forces,
  referenced to the ambient wind $v_{\rm w}$ and the tether force
  $F_{\rm w}$ corresponding to an air path speed of $v_{\rm w}$.
  At first sight it is counter-intuitive, that increasing negative winch speeds,
  while winching in, lead to decreasing tether forces for constant glide ratios.
  It should be emphasized, that the presented cycle scheme explicitly exploits this feature
  during the return phase in order to improve the cycle overall power
  generation efficiency. However, it is worth mentioning the drawback, that 
  negative winch speeds can lead to critical unstable flight situations, if
  e.g.~the winch is stopped suddenly, compare end of Sect.~\ref{sec:plant_winching}.}
  \label{fig:winching}
\end{figure}

In order to examine the effect of winching, the steady state for constant
winching speed $\dot{l}$ is calculated by setting $\dot{\vartheta}=0$ in
(\ref{eq:eqm_theta2}). Some trigonometric manipulations yield for the
equilibrium wind window angle
\begin{equation}
  \vartheta_0^{\rm (winch)} = \vartheta_0 - \arcsin \left[
  \frac{\cos\psi}{\sqrt{\cos^2\psi + (1/E)^2}}\frac{\dot{l}}{v_{\rm w}}
  \right]
  \label{eq:theta_winching}
\end{equation}
with the zenith position for zero winch speed of
$\vartheta_0=\arctan(E\cos\psi)$.
The corresponding curve is shown in
Fig.~\ref{fig:winching}.

Note, that reeling in at windward position in order to decrease the tether force
is explicitly exploited in the cycle scheme presented here.

\begin{figure}
  \centering
  \includegraphics[width=8.8cm]{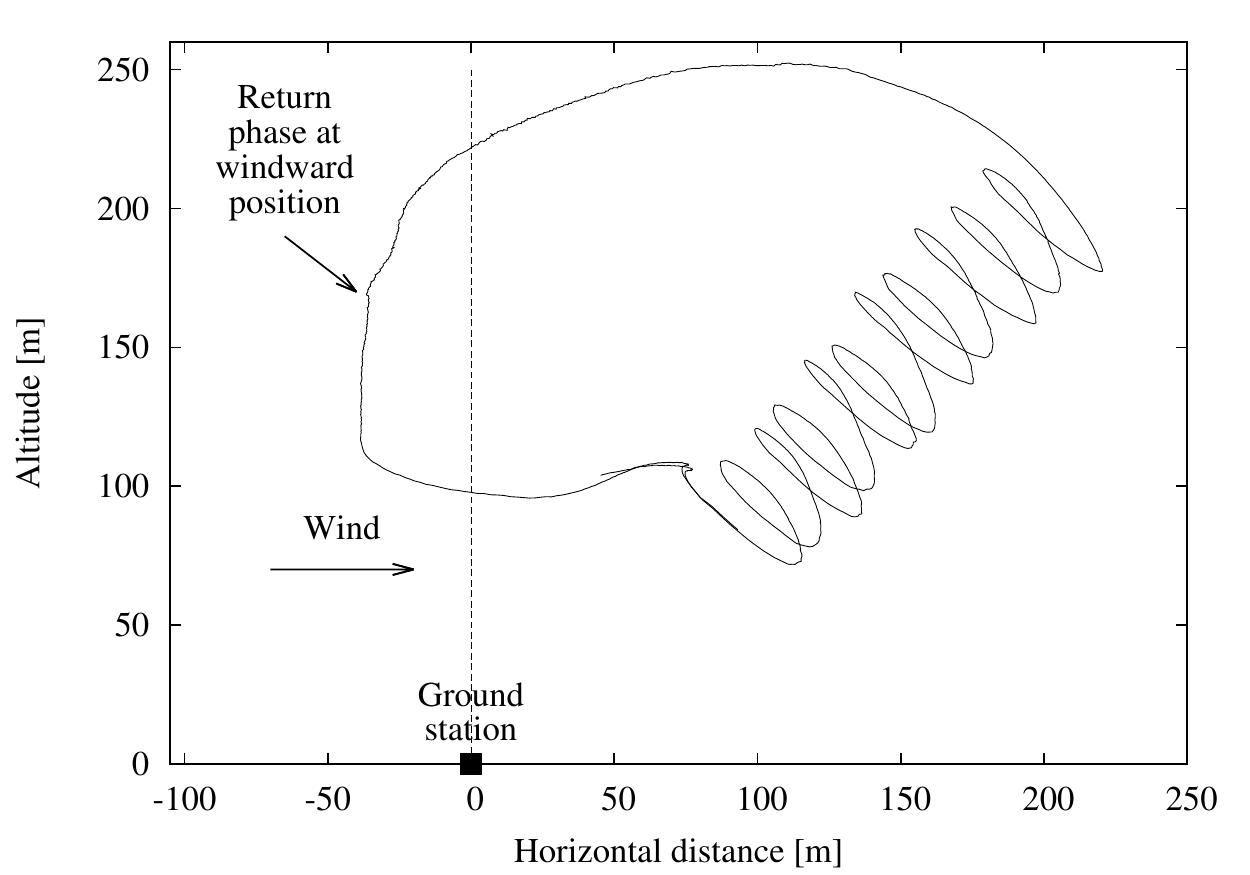}
  \caption{Side view of trajectory for one power cycle. Note, that during the
  return phase, the kite is flown in a windward position in order to further
  reduce the tether force, compare Fig.~\ref{fig:winching}.}
  \label{fig:plot_side_view}
\end{figure}

In order to further illustrate this effect of winching, the side view of a cycle
trajectory is plotted in Fig.~\ref{fig:plot_side_view}.
Finally, a brief
comment on the winching-in phase shall be given. In order to keep the
kite in a stable flight configuration, a certain minimum air path speed
$v_{\rm a,min}$ is required.
Assuming $v_{\rm a}>v_{\rm a,min}$ and resolving (\ref{eq:eqm_va})
w.r.t.~$\dot{l}$ yields:
\begin{equation}
  \dot{l} < -\frac{v_{\rm a,min}}{E} + v_{\rm w}\cos\vartheta
  \label{eq:dot_l_min}
\end{equation}
It can be seen, that for angles $\vartheta>\pi/2$ a certain winch speed
$\dot{l}<0$ is needed in order to keep the stable flight configuration. Before
stopping the winch, the angle $\vartheta$ has to be reduced by flying into
the wind window as is done by starting the power phase. This issue imposes
the reliability requirement of 'avoiding sudden stops by all means' to the winch
setup.

%% file: control_overview.tex
\section{Control design overview}
\label{sec:control_overview}
In this section, a brief overview on the complete control system is
given. Details of the controller parts and presentation of experimental results
will be given in the subsequent sections.
\subsection{Design philosophy}
Before diving into the details of the control system, the design ideas and
principles are summarized in order to provide a kind of justification for the
described setup in the rest of the paper.
Examining the equations of motion (\ref{eq:eqm_psi})--(\ref{eq:eqm_l}),
a possible control approach could be based on the concept of dynamic inversion
or model predictive control \cite{Lucia2014a}.
In the field of tethered kite control, such approaches have been applied to 
much more complex theoretical models \cite{Williams2008a}, \cite{Gros2014a} than
the model of Sect.~\ref{sec:plant}, and the principal functionality has been 
demonstrated successfully in simulations.
However, successful applications of these involved algorithms on real prototypes has not
been reported yet.

A prototype setup and the task of developing
an operationally robust, commercial control system imposes a different emphasis and leads
to a different controller structure.
The plant is subject to huge perturbations due to wind gusts  with
uncertainties about wind conditions over the range of flight altitudes even for one
cycle.
In order to successfully tackle these real world and industrial challenges, 
one prefers certain control topologies and design principles,
which will be summarized in the following:
\begin{itemize}
  \item An important step is to split up the system into separate parts which can be
described, to a large degree, by analytic equations, related to intuitive physical models. 
For each subsystem, it is much easier to design robust, but still simple and
linear controllers, augmented by nonlinear elements like limiters, which can be
easily inserted and tuned.
This approach naturally leads to a cascaded controller topology. 
  \item Implement a feedforward/feedback structure in order to achieve the bandwidth
  needed for tailored curve flight and to capture the major non-linearities in the feedforward
  path. 
  The feedforward paths also allow for a proper
  shaping of signals according to system constraints and can be easily added to
  a cascaded design.
  \item  Modify the open loop dynamics by the feedback to the necessary degree only.
  As a consequence, we do not try to follow an
  exact predefined trajectory at all cost. The achievable  control bandwidth of the inner loop
  would not allow for very high accuracy, when taking robustness as major design consideration
  into account. Instead, the simple target point
  tracking scheme proposed in \cite{Fagiano2013a} was implemented, which
  supports a kind of natural evolution of the eight-pattern, but keeps the
  pattern reliably in the desired region of the wind window on the other hand.
\end{itemize}

\begin{figure}
  \centering
  \includegraphics[width=8.8cm]{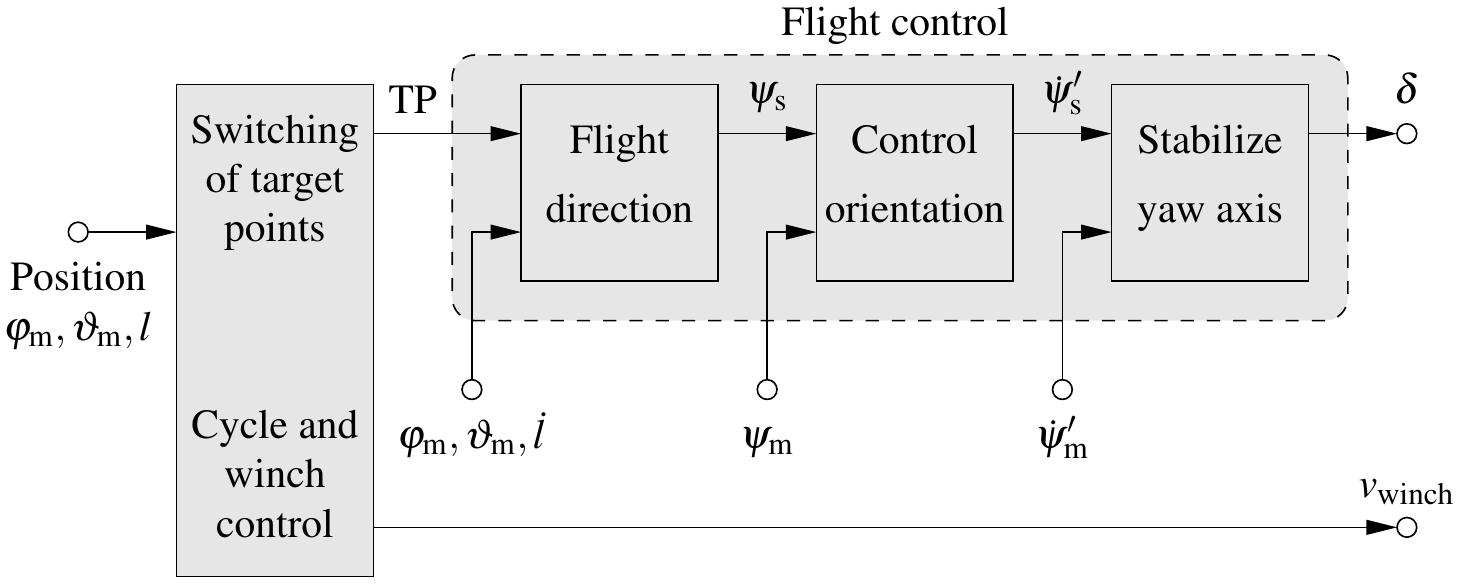}
  \caption{Overview on the control system.}
  \label{fig:control_overview}
\end{figure}

The implemented overall control system structure is depicted in
Fig.~\ref{fig:control_overview}.
The control strategy is based on flight control towards target points
(TPs), while switching to a subsequent target point before the
current target point is reached \cite{Fagiano2013a}. This
algorithm, which also controls the winch, is called {\em winch and cycle
control}. Compared to \cite{Fagiano2013a}, where a simple heading control
has been implemented, we present a cascaded {\em flight control} setup
based on earlier work \cite{Erhard2012a}, providing accurate heading control as well
as shaped curve flight, which are both prerequisites for robust eight-down
flight.
\subsection{Winch and cycle control}
This part is responsible for the overall control of the power generation cycle
and the computation of the winch speed set value $v_{\rm winch}$.
\begin{enumerate}
  \item The {\em cycle control} in its simplest,
  albeit not fully optimized, version can be based on three target
  points only. The geometry of these target points is indicated in
  Fig.~\ref{fig:target_points}. Switching between target points is triggered by
  geometric conditions, i.e.~by approaching condition to the active target point
  as well as based on the line length $l$ in order to obtain the
  periodic repetitions of cycles consisting of power and return phases. Details
  will be given in Sect.~\ref{sec:guidance}.
  \item In order to implement an efficient {\em winch control} algorithm, the
  overall optimization problem involving complete cycles has to be tackled as suggested
  in \cite{Horn2013a}. However, as wind conditions in different flight altitudes
  are not exactly known and subject to gusts and significant variations, simpler
  approaches are desirable at least for first proof-of-concept flight tests.
  Performing a rudimentary numerical optimization with the model
  (\ref{eq:eqm_psi}--\ref{eq:eqm_va}), a simple relation was
  identified, which allows for the computation of the winch speed based on the
  geometric condition of wind window position and wind speed only. Although this approach
  is one of the simplest, it is performing surprisingly well as it
  is capable of operating all phases and the energy production seems not to
  be much away from the achievable optimum. Details will be given in
  Sect.~\ref{sec:control_winch}.
\end{enumerate}
\subsection{Flight control}
The flight control design is composed of three cascaded controllers as shown
in Fig.~\ref{fig:control_overview}. In the following, these blocks are
described from right to left.
\begin{enumerate}
  \item Inner loop ($\dot{\psi}^\prime$-controller): Yaw axis stabilization with set
  point turn rate $\dot{\psi}_{\rm s}^\prime$. The plant behavior $\delta \rightarrow
  \dot{\psi}^\prime$ is based on the turn rate law
  \begin{equation}
     \dot{\psi}^\prime = g_{\rm k}\, v_{\rm a}\, \delta
     \label{eq:trl_inner_loop}
  \end{equation}
  Note, that the inertial turn-rate sensor (yaw axis) measures the turn
  rate $\dot{\psi}_{\rm m}^{\prime}$ which corresponds to $\dot{\psi}^\prime$
  and not $\dot{\psi}$.
  
  \item Outer loop ($\psi$-controller): $\psi_{\rm
  s}$ angle control (related to flight direction). The plant behavior $\dot{\psi}_{\rm s}^\prime \rightarrow \psi$ is given by
  \begin{equation}
     \dot{\psi} = \dot{\psi}_{\rm s}^\prime + \dot{\psi}_{\rm ct} =
     \dot{\psi}_{\rm s}^\prime + \dot{\varphi}\cos\vartheta 
     \label{eq:plant_psi}
  \end{equation}
  It should be remarked, that the controller deals with multiple input
  values $[\psi_{\rm s}, \dot{\psi}_{\rm ct}] \rightarrow \dot{\psi}_{\rm
  s}^\prime$ where $\dot{\psi}_{\rm ct}$ involves further quantities given by
  (\ref{eq:crossterm_operational}).
  However, it is appropriate to assume $\dot{\psi} \approx \dot{\psi}^\prime$
  for the initial design and add the crossterm as compensating correction.

  \item Guidance: Control flight towards target point. The flight
  direction is based on wind window position, hence the guidance computes
  \begin{equation}
     [\varphi_{\rm m},\vartheta_{\rm m},
      \varphi_{\rm TP},\vartheta_{\rm TP}] \rightarrow \psi_{\rm s}
  \end{equation}
  The guidance is done by computing the flight direction $\gamma$ towards a
  target point and subsequently $\psi_{\rm s}$ by inverting
  (\ref{eq:flight_direction}). It should be finally noted, that due to the
  switching of target points, discontinuities are imposed on $\psi_{\rm s}$. As
  these steps $\psi_{\rm s}$ should result in well defined and controlled curve
  flights, a proper shaping has to be performed. In the presented design
  this is done in the $\psi$-controller as described in
  Sect.~\ref{sec:control_psi} in detail.
\end{enumerate}

%% file: control_psidot.tex
\section{Controller for yaw rate $\dot{\psi}^\prime$}
\label{sec:control_psidot}
\begin{figure*}
  \centering
  \includegraphics[width=\textwidth]{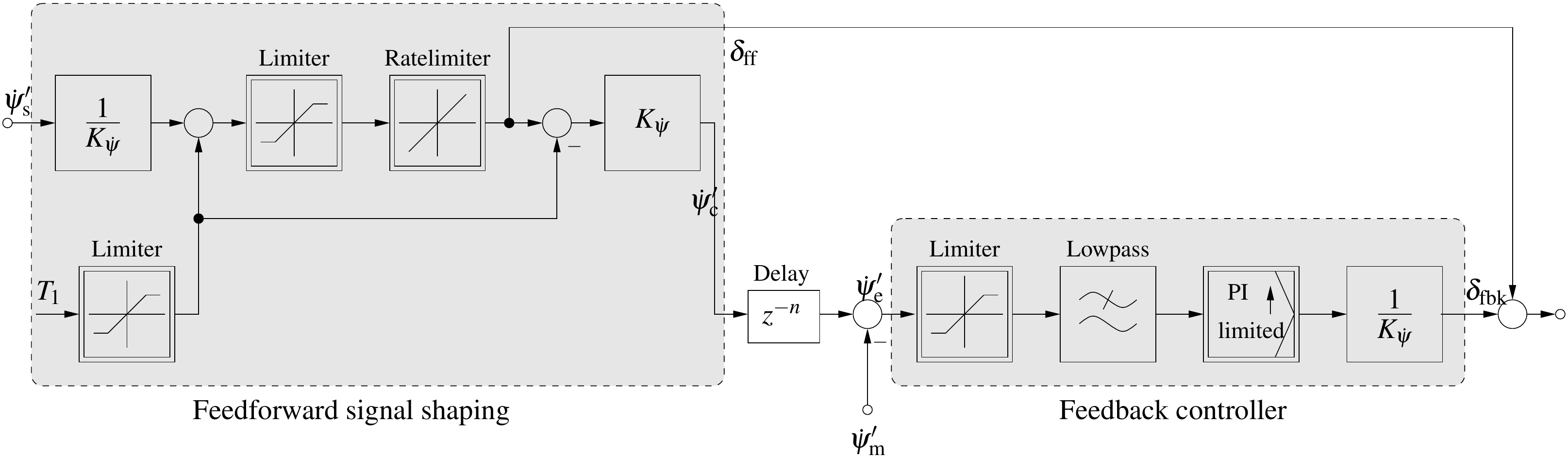}
  \caption{Setup of the $\dot{\psi}$-controller based on a feedforward and
  feedback structure. The limiter and rate limiter in the feedforward shape
  the signal according to limited steering range and speed of the
  actuator in the control pod. Note the $1/K_{\dot{\psi}}$-block in order to
  obtain a linear plant behavior.}
  \label{fig:psidot_controller}
\end{figure*}

The complete setup of the $\dot{\psi}^\prime$-controller is shown in
Fig.~\ref{fig:psidot_controller}.
The controller is based on the feedforward and feedback parts marked by the
dashed gray boxes. 
The plant behavior is based on the turn-rate law, compare
(\ref{eq:trl_inner_loop})
\begin{equation}
  \dot{\psi}^\prime = g_{\rm k}\, v_{\rm a}\, \delta \label{eq:ctl_psidot_trl}
\end{equation}
In order to obtain a linear plant behavior with stationary parameters, the
dependence on $v_{\rm a}$ is eliminated by the $1/K_{\dot{\psi}}$ block in the feedback controller with 
$K_{\dot{\psi}}=g_{\rm k} v_{\rm a}$.
This creation of a meta-actuator is based on the assumption, that
$K_{\dot{\psi}}$ changes slowly compared to the $\dot{\psi}^\prime$ dynamics.
Due to the proportional nature of the linearized plant 
($\dot{\psi}^\prime =  K_{\dot{\psi}} \delta$), a PI-controller is
used for the control task in addition with a low-pass to suppress unwanted frequency
components. Further, a limiter is applied to the error signal $\psi_{\rm e}$ for
safety reasons. It should be mentioned here, that the source of the turn rate
measurement $\dot{\psi}_{\rm m}$ is a single inertial turn-rate sensor aligned
in yaw axis of the flying system.

The feedforward command is computed based on
$\delta_{\rm ff}=\dot{\psi}_{\rm s}$ and shaped by a limiter and rate limiter,
in order to take into account the limited deflection range and steering
speed of the control pod, respectively.
It should be mentioned, that gravity leads to an additional term in the
turn-rate law, which reads \cite{Erhard2012a, Erhard2013b}
\begin{equation}
  \dot{\psi}^\prime = K_{\dot{\psi}} \,\delta + M\frac{\cos\theta_{\rm g}
  \sin\psi_{\rm g}}{v_{\rm a}} \label{eq:extended_trl}
\end{equation}
with a system weight-dependent parameter $M$. Note, that the angles 
$\theta_{\rm g}$ and $\psi_{\rm g}$ are defined with respect to a different
coordinate system. For details on definition and origin, the reader is kindly
referred to \cite{Erhard2012a, Erhard2013b}. 
In order to compensate for the gravity term, the quantity
\begin{equation}
  T_1 = \frac{M}{K_{\dot{\psi}}}\frac{\cos\theta_{\rm g}
  \sin\psi_{\rm g}}{v_{\rm a}}
\end{equation}
can be fed into the feedforward block. It should be noted, that the $T_1$
input feature is given for sake of completeness only. For usual operating
conditions, the gravity compensation could be switched off ($T_1=0$) and the neglected effect in the feedforward path is
easily dealt with by the feedback path. In addition, as $M$ depends on the weight
of the flying system, a dependence on $l$ should be taken into account. The accurate compensation of
this term, taking into account varying line lengths, is subject of current research
and will be published elsewhere.

%% file: control_psi.tex
\section{Controller for yaw angle $\psi$}
\label{sec:control_psi}
\begin{figure*}
  \centering
  \includegraphics[width=\textwidth]{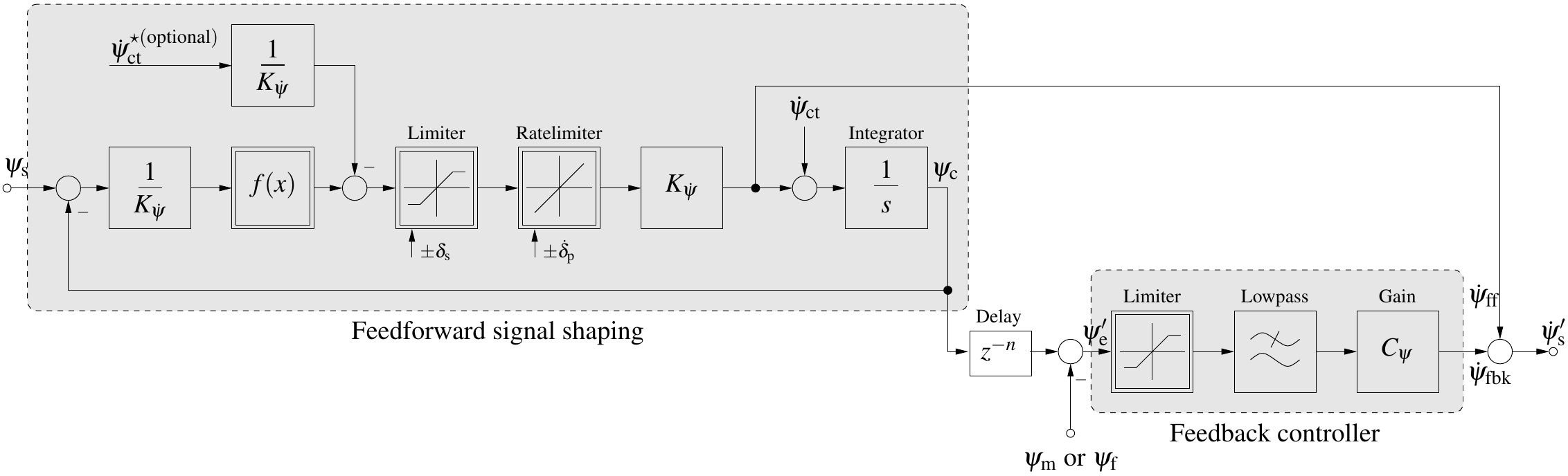}
  \caption{Setup of the psi-controller as feedforward / feedback parts.
  The feedforward block features an internal loop and is capable of shaping
  step-wise $\psi_{\rm s}$ input signals taking into account the
  given limitations of steering range and speed of the control pod actuator.}
  \label{fig:psi_controller}
\end{figure*}

The complete setup of the $\psi$-controller is given in
Fig.~\ref{fig:psi_controller}.
Like the inner loop controller described in the previous section, it features a
feedforward/feedback structure. As the plant dynamics, apart from the crossterm
correction, is of integrator type, the feedback is implemented as a 
proportional controller with
preceding low pass filter and limiter for safety reasons.

The feedforward part is more involved, since it has to meet the following
requirements. As already introduced, the switching of target points imposes steps on
the set value $\psi_{\rm s}$. These step discontinuities are related to
commanded changes of flight direction and thus have to be implemented as
properly curved flights, which should take into account limits on steering
deflection $\delta_{\rm s}$ and steering speed $\dot{\delta}_{\rm p}$. The
resulting flown curve radii must be small enough for efficiency reasons and
for a safe fitting into a limited space of the wind window. In
addition, the crossterm correction $\dot{\psi}_{\rm ct}$ has to be considered
according to (\ref{eq:plant_psi}). 
The shaping of $\psi_{\rm s}$ could be done by a
low-pass filter \cite{Fagiano2013a}. However, 
in order to get time-optimal trajectories for $\psi_{\rm s}$ meeting the
given constranints, an internal loop is implemented, which will be discussed in the
following.

The control pod is modeled by a limiter and a rate limiter for steering
deflection limit $\delta_{\rm s}$ and steering speed $\dot{\delta}_{\rm p}$,
respectively. 
The integrator implements $\psi_{\rm c}=\int dt\,(\dot{\psi}_{\rm ff} +
\dot{\psi}_{\rm ct})$, which reflects the plant dynamics, compare to
(\ref{eq:plant_psi}). The scaling from rates to steering and back is
done by the $1/K_{\dot{\psi}}$ and $K_{\dot{\psi}}$ blocks, respectively.
The feedback scaling function $f(x)$ implements the inverse of the modeled plant
in order to achieve time optimal following of $\psi_{\rm c}\rightarrow \psi_{\rm
s}$. In order to determine $f(x)$, the process starting with an initial
deflection $\delta_{\rm i}>0$ and steering with speed $-\dot{\delta}_{\rm p}$ to a target
deflection $\delta_{\rm t}$ is considered.
At target deflection, attainment of set point is assumed, i.e.~$\psi_{\rm
c}=\psi_{\rm s}$ and $\dot{\psi}_{\rm c}=\dot{\psi}_{\rm s}$. The latter implies
\begin{equation}
  K_{\dot{\psi}} \delta_{\rm t} + \dot{\psi}_{\rm ct} = \dot{\psi}_{\rm s}
\end{equation}
Inserting the steering $\delta_{\rm t}=\delta_{\rm i}-\dot{\delta}_{\rm p}t$ and
resolving w.r.t.~$t$ yields:
\begin{equation}
  t = \frac{\delta_{\rm i}-(\dot{\psi}_{\rm s}-\dot{\psi}_{\rm
  ct})/K_{\dot{\psi}}}{\dot{\delta}_{\rm p}}
  \label{eq:t_process}
\end{equation}
The corresponding $\Delta \psi$ of this process can be computed as
\begin{equation}
  \Delta \psi = \int\limits_0^t dt^\prime K_{\dot{\psi}} (\delta_{\rm
  i}-\dot{\delta}_{\rm p}t^\prime)\label{eq:ff_block_delta_psi}
\end{equation}
and inserting (\ref{eq:t_process}) yields
\begin{equation}
  \Delta\psi = \frac{(\delta_{\rm i}+\dot{\psi}_{\rm
  ct}/K_{\dot{\psi}})^2}{2\dot{\delta}_{\rm p}}
\end{equation}
This $\Delta\psi$ is interpreted as error, resolving this equation
w.r.t.~$\delta_{\rm i}$ and reads
\begin{equation}
  \delta_{\rm i} = \sqrt{\frac{2\dot{\delta}_{\rm
  p}\,\Delta\psi}{K_{\dot{\psi}}}} - 
  \frac{(\dot{\psi}_{\rm ct} - \dot{\psi}_{\rm s})}{K_{\dot{\psi}}}
  \label{eq:delta_i}
\end{equation}
As $\delta_{\rm i}$ is 
deflection related to this error, it is an
appropriate feedback value.
Comparing (\ref{eq:delta_i}) to the diagram and generalizing by consideration of
$\delta_{\rm i}<0$ results in the following
\begin{equation}
    f(x)= {\rm sign}(x) \sqrt{2\dot{\delta}_{\rm p}|x|}
\end{equation}
For constant or step-wise $\psi_{\rm s}$ input signals, one would chose
\begin{equation}
  \dot{\psi}_{\rm ct}^\star=\dot{\psi}_{\rm ct} \label{eq:ct_star_input}
\end{equation}
For $\psi_{\rm s}$ inputs based on target points, $\dot{\psi}_{\rm ct}^\star=0$
is the appropriate choice as can be reasoned as follows: when heading to a target point
and the final course is reached, the motion could be regarded approximately as free
'inertial' motion, which implies, compare (\ref{eq:psidot_ct}) and text below,
$\dot{\psi}_{\rm s}\approx \dot{\psi}_{\rm ct}$. Inserting into (\ref{eq:ct_star_input}) implies
$\dot{\psi}_{\rm ct}^\star=0$.

%% file: control_targetpoint.tex
\section{Flight direction control}
\label{sec:target_point}
\begin{figure}
  \centering
  \includegraphics[width=6cm]{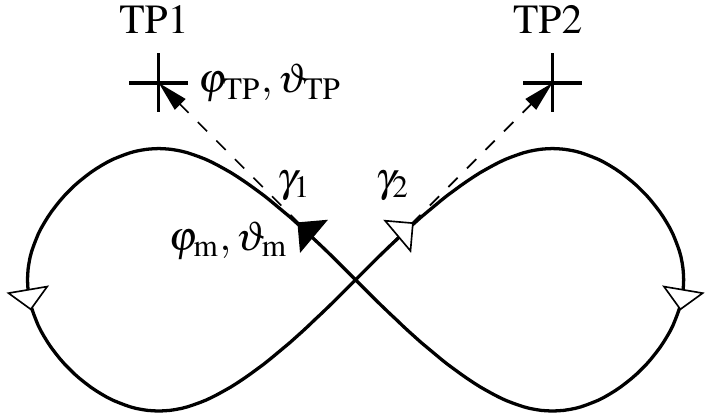}
  \caption{Pattern generation by heading towards target points. Clockwise
  change of direction in this view is defined as increase of $\gamma$.}
  \label{fig:target_points}
\end{figure}

As already introduced, pattern generation is accomplished by navigating
towards target points.
The basic principle is sketched in Fig.~\ref{fig:target_points}.
Based on the great-circle navigation, which determines the shortest
connection between two given points on the sphere, the direction from the
current position $\varphi_{\rm m},\vartheta_{\rm m}$ to the target point $\varphi_{\rm
TP},\vartheta_{\rm TP}$ can be computed by 
\begin{eqnarray}
    \gamma_{\rm s} &=& \arctan \left(
    \sin(\varphi_{\rm m}-\varphi_{\rm TP}),\right. \\
    &&\left.\quad\quad
    \cos\vartheta_{\rm m}\cos(\varphi_{\rm TP}-\varphi_{\rm m}) - 
    \cot\vartheta_{\rm TP} \sin\vartheta_{\rm m}\right)\nonumber
    \label{eq:direction}
\end{eqnarray}
Having determined the flight direction $\gamma_{\rm s}$, the set value for
$\psi_{\rm s}$ is computed by inversion of (\ref{eq:psi_from_gamma}). 

In order to perform a curve flight, the target point is switched, as from TP1
to TP2 in this example. This switching leads to a step in $\psi_{\rm s}$.
However, due to the shaping in the $\psi$-feedforward as explained in
Sect.~\ref{sec:control_psi}, a smooth and well-controlled curve will be
commanded. The nominal steering deflection value $\delta_{\rm s}$ determines the
radius of the curve, which can be estimated by comparing the approximation
for the tangential speed $\dot{\psi}_{\rm m}^\prime \approx v_{\rm
a}/r_{\rm curve}$ with (\ref{eq:ctl_psidot_trl})
\begin{equation}
  r_{\rm curve} = \frac{1}{g_{\rm k} \delta_{\rm s}}
\end{equation}

Finally, the issue of unwrapping the course angles shall be explained. Using
(\ref{eq:direction}) one would obtain for the directions in Fig.~\ref{fig:target_points} 
e.g.~$\gamma_1=1.0$\,rad and $\gamma_2=-1.0$\,rad, respectively. However, the
curve $\gamma = 1.0 \rightarrow (-1.0)$ would be clockwise and not
counter-clockwise as needed for the drawn figure-eight.
Hence $\gamma_2=(2\pi-1.0)$ has to be chosen.
Alternatively, the same figure could be parameterized by $\gamma_1=(-2\pi+1.0)$
and $\gamma_2=-1.0$. The modulo-$2\pi$ offset could be freely chosen as initial
condition, but has to be kept constant during pattern operation.

%% file: control_cycle.tex
\section{Cycle control}
\label{sec:guidance}
\begin{figure}
  \centering
  \includegraphics[width=6cm]{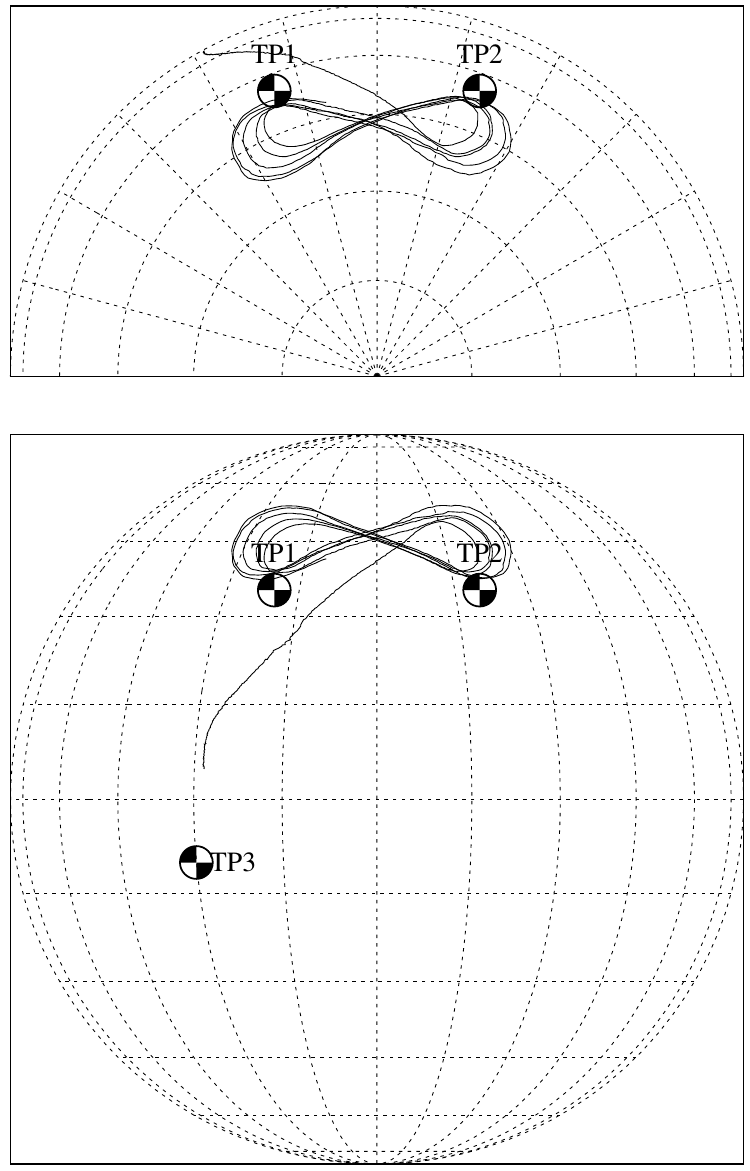}
  \caption{Projection of the flight trajectory on the unit sphere, horizontal
  view in wind direction (upper figure) and top view (lower figure). Shown are power and
  transfer phase. Note, that the varying tether length is not represented here.
  The indicated coordinate grids correspond to a spherical coordinate system
  with symmetry axis aligned in wind direction.}
  \label{fig:plot_angles}
\end{figure}

In this section, the flight control generating the pattern-eight as well as
steering the kite during the return phase is presented. 
In order to illustrate the target point method, projections of the flight
trajectory on the unit sphere including target points are shown in
Fig.~\ref{fig:plot_angles}.
 
\begin{figure}
  \centering
  \includegraphics[width=6cm]{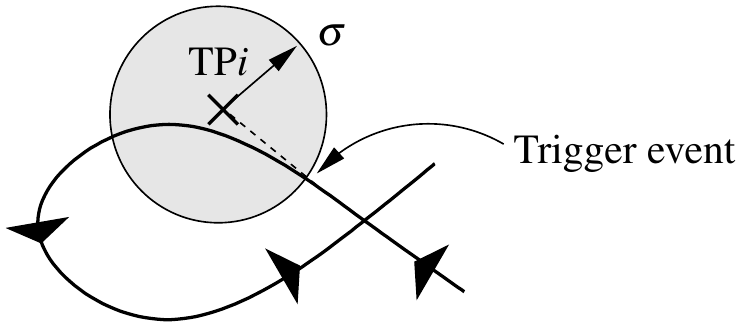}
  \caption{Trigger condition. The active target point is switched to a
  subsequent target point, when the 'angular' distance on the unit sphere
  drops below a certain threshold as indicated by the gray circle.}
  \label{fig:trigger}
\end{figure}

The general principle of the target points is basically to control the flight
direction heading towards an active target point. 
The dynamic pattern is generated by switching to another target point before the
currently active target point is reached. 
The eight-down for the power phase is guided by the two target points TP1 and
TP2, compare Fig.~\ref{fig:plot_angles}, with the coordinates chosen
symmetrically with respect to the vertical axis as follows:
$\varphi_{\rm TP2}=-\varphi_{\rm TP1}$ and $\vartheta_{\rm TP2}=\vartheta_{\rm
TP1}$.
The trigger condition for switching to the subsequent target point is
defined with respect to the 'angular' distance on the unit sphere as follows:
\begin{equation}
  (\varphi_{\rm m}-\varphi_{{\rm TP}i})^2 \sin^2 \vartheta_{{\rm TP}i} + 
  (\vartheta_{\rm m}-\vartheta_{{\rm TP}i})^2 \leq \sigma^2 
\end{equation}
This condition is graphically illustrated in Fig.~\ref{fig:trigger}.
The value $\sigma$ has been chosen empirically and kept constant for the
results presented here. However, in order to optimize the pattern, a dependence
$\sigma=\sigma(l)$ could be introduced.

For the transfer and return phases, the kite is flown towards target point TP3,
compare Fig.~\ref{fig:plot_angles}. In contrast to the power phase, no switching is
performed during these phases. In order to get a reasonable feedback, the
target point should not be chosen too far away. In addition, it has to be
made sure, that the target point is never reached, since the flight direction
would become undefined due to the singularity there, compare
(\ref{eq:direction}).
Hence its elevation value is chosen dynamically dependent on the current position
\begin{equation}
  \vartheta_{\rm TP3}= {\rm max} (\pi/2,\vartheta_{\rm m}+\Delta\vartheta)
\end{equation}
with a typical value $\Delta \vartheta = 0.3$\,rad.
The azimuth coordinate has to be chosen as a compromise. For 
$\varphi_{\rm TP3}=0$, no influence of gravity on the
steering behavior would be present, compare (\ref{eq:extended_trl}), but
the tether tension would be maximally reduced by gravity.
For values $\varphi_{\rm TP3}>0$, the influence of gravity on tether
force is reduced allowing for lower operational wind speeds, but the effect of gravity on
steering increases and sufficient space above the surface has to be left for the
maneuvers.
As a consequence, the chosen value is typically $\varphi_{\rm
TP3}=0.4$\,rad.

\begin{figure}
  \centering
  \includegraphics[width=7cm]{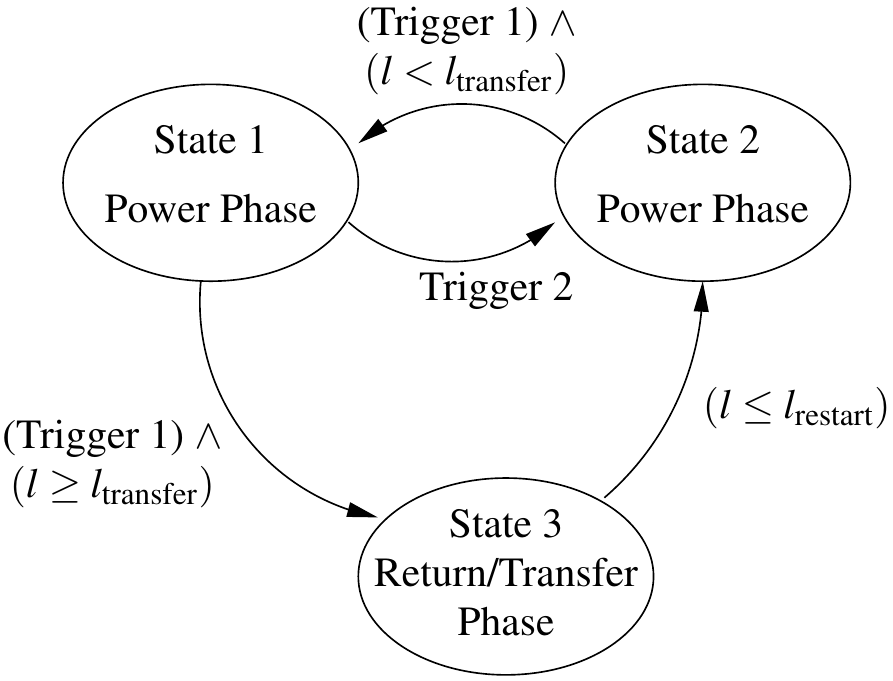}
  \caption{State diagram for the overall cycle control. During the power
  phase, switching between the target points TP1 and TP2 is based on geometrical
  trigger conditions, compare Fig.~\ref{fig:trigger}. 
  Activation of the Transfer phase is triggered above a maximal line
  length $l_{\rm transfer}$, in addition to the geometric condition. The return
  phase is ended below $l_{\rm restart}$.}
  \label{fig:states}
\end{figure}

The complete state diagram for the overall cycle control is shown in
Fig.~\ref{fig:states}.
Note, that during the power phase, switching of target points is triggered by
geometrical conditions while begin and end of the transfer/return-phases are
determined by line length limits, which are in this case 
$l_{\rm restart}=130$\,m and $l_{\rm transfer}=270$\,m. 
Finally, it should be remarked that the transition from
states 1 to 3 is due to the assumption $\varphi_{\rm TP3}>0$. The according
diagram for negative values $\varphi_{\rm TP3}<0$ follows straight forward from
symmetry considerations.

%% file: control_winch.tex
\section{Winch control and power generation}
\label{sec:control_winch}
First it should be remarked, that the setup of an electrical motor/generator
attached to a frequency converter involves internal control loops, which are not
subject to this paper as they have been tuned according to the respective user
manuals.
However, one feature is worth mentioning. The current control loop of the
frequency converter is used to limit the maximal tether load and avoid
overload of the kite by setting the maximal current accordingly.
This is a very effective mechanism, as the current control loop is very fast and
therefore, apart from inertia effects of the moving elements of winch and
motor/generator, depowering is done as fast as possible.

The following section focuses on computing the set value for the winch
speed $v_{\rm winch}$ in order to operate efficient power cycles.

It should first be noted, that a computation of the winch speed for
optimal power output calls for solving an optimization problem
considering complete power cycles, which is quite involved and subject to current research
activities \cite{Costello2013a}, \cite{Horn2013a}.
However, these extended models are far beyond the scope of getting a
small-scale prototype setup operational in order to prove, that
fully automatic power generation is feasible. 
Hence, simpler approaches are needed, which compute the winch speed based on the
system state as e.g.~given in \cite{fechner2014feed}, which proposes a
feedforward implementation for constant force.

The main idea is to separate flight and
winch control on the kinematic level. This is accomplished by flying the
pattern geometrically, e.g.~guided by target points defined on the unit sphere as
primary control and commanding the winch speed dependent on the current pattern.
This is also the control strategy chosen when operating the prototype by two
human operators. The pilot flies the pattern-eight and static positions,
respectively, while the winch operator commands the winch speed accordingly.
This task comprises basically winching out during the pattern and
winching in during the static flight position, respectively. Hence, the goal is
to find simple controllers for the winch speed for the different phases, which will
be considered separately in the following.
It has to be remarked, that for sake of clarity, only the basic functionalities
are given, which were used for the presented experimental results. They may lack
robustness w.r.t.~untypical wind situations or temporary free (quasi untethered, i.e. a non stretched line)
flight. Operational extensions are subject to current development and would go
far beyond the scope of this paper.
%
%
\subsection{Power phase winch control}
\label{sec:power_phase_winch_control}
For the power phase, the thumb-rule using 1/3 of the
projected wind speed could be used \cite{Loyd1980,Fagiano2012a}. 
Further optimizations have been proposed \cite{Luchsinger2013a}, which suggest
a different factor $(1/a)<(1/3)$ by taking into account the return phase.
Thus the set point for the winch speed is given by
\begin{equation}
  \dot{l}_{\rm s} = \frac{1}{a} v_{\rm w}^\prime \cos\vartheta
  \label{eq:ldot_s}
\end{equation}
The $v_{\rm w}^\prime$ value denotes the wind speed at flight position. Note,
that $v_{\rm w}^\prime$ is hardly measurable directly.
However making use of (\ref{eq:eqm_va}), which could be regarded as kind of
plant $\dot{l} \mapsto v_{\rm a}$ with
\begin{equation}
  v_{\rm a} = v_{\rm w}^\prime E \cos\vartheta - \dot{l}_{\rm s}
  E\label{eq:eqm_va2}
\end{equation}
and the simple proportional feedback 
\begin{equation} 
  \dot{l}_{\rm s}=\frac{v_{\rm a}}{(a-1)E}
\end{equation}
it can be shown, that this simple loop fulfills the requirement
(\ref{eq:ldot_s}) as the stationary value reads $\dot{l}\rightarrow v_{\rm
w}^\prime (\cos\vartheta)/a$.

\begin{figure}
  \centering
  \includegraphics[width=8.8cm]{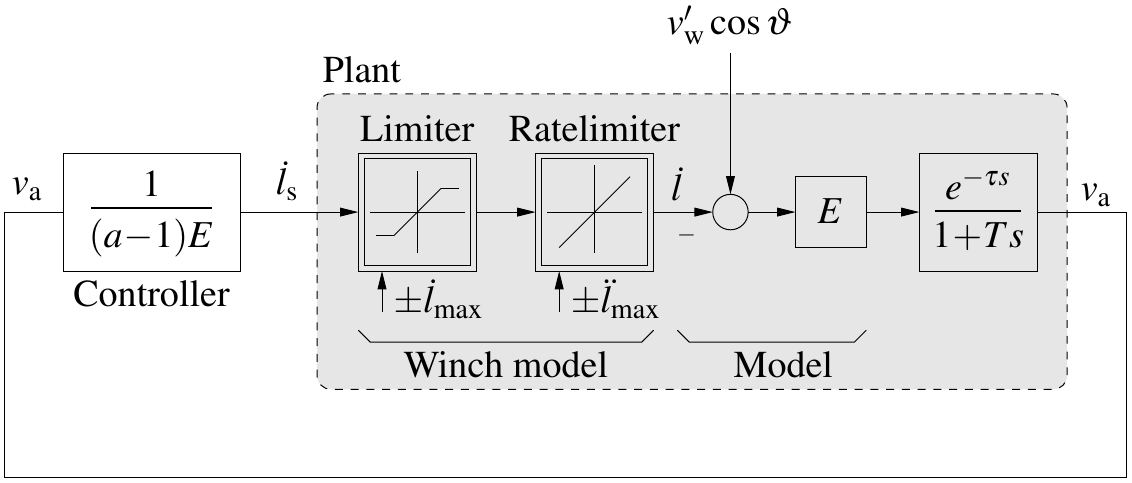}
  \caption{Controller for the power phase based on a proportional feedback of
  the air path velocity. The wind speed $v_{\rm w}^\prime$ is the back wind at
  the flight position.}
  \label{fig:winch_feedback}
\end{figure}

The setup is drawn schematically in Fig.~\ref{fig:winch_feedback}.
In order to allow for a realistic simulation, the main practical constraints of
limited speed $\dot{l}_{\rm max}$ and acceleration $\ddot{l}_{\rm max}$ of the
winch as well as delay $\tau$ and low pass behavior of the system have been
added to the figure, but will not be further discussed here.
Note, that due to the loop setup, gusts on $v_{\rm w}^\prime$ are anticipated by
$\dot{l}$ leading to an efficient and consistent behavior of the winch during
the power phase. 

It should be finally remarked, that 
winch torque could be used as control variable instead of winch speed
by applying a torque set value proportional to the square of the measured winch
speed. This leads to quite similar results
\cite{Zgraggen2014b} and could be used alternatively. In
addition, winch inertia might play a non-negligible role on system dynamics.
The optimization of the winch control system with respect to efficient power
cycles is subject to current research activities. 
\subsection{Transfer and return phase winch control}
In view of overall efficiency, winching in as fast as possible at tether
forces as low as possible would be desirable. 
In order to achieve this, {\em tuning} the glide ratio has been proposed and 
performance modeling for such systems has been presented
\cite{Luchsinger2013a}, \cite{Fechner2013a}.
In contrast, our system is operated at {\em constant} glide
ratio, but also allows for low-force return phases as discussed
in Fig.~\ref{fig:winching}.
In order to understand how to choose the winch speed for
efficient transfer and return phases, an optimization problem based on the
model of Sect.~\ref{sec:plant} has been solved. 

\begin{figure}[h]
  \centering
  \includegraphics[width=8.8cm]{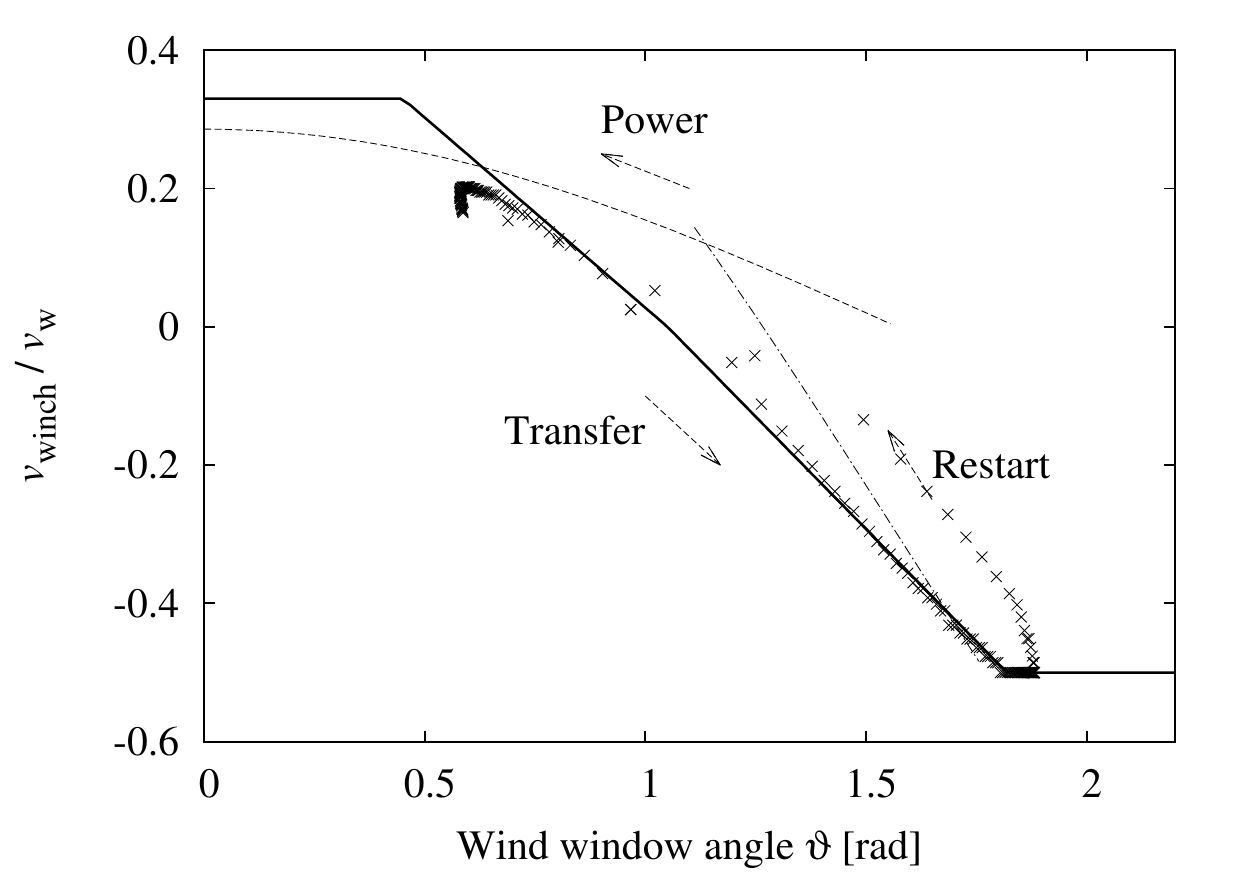}
  \caption{Winch speed as function of windwindow angle. Simulation results of
  optimized cycles are plotted as points, compare the two middle subplots of
  Fig.~\ref{fig:plot_simulation}.
  The data points of the transfer phase could be fitted by a linear saturated
  transfer function given in Fig.~\ref{fig:winch_controller} and shown as 
  solid line.
  For sake of completeness, outputs of the power phase winch controller
  for $a=3.5$ (dotted line) and the restart winch controller for $v_{\rm
  min}=1.5 v_{\rm w}$ (dash-dotted line) are given (compare
  Sect.~\ref{sec:power_phase_winch_control} and
  Sect.~\ref{sec:restart_control}).
  The whole cycle is operated as indicated by the arrows.}
  \label{fig:winch_function}
\end{figure}

\begin{figure}[h]
  \centering
  \includegraphics[width=7cm]{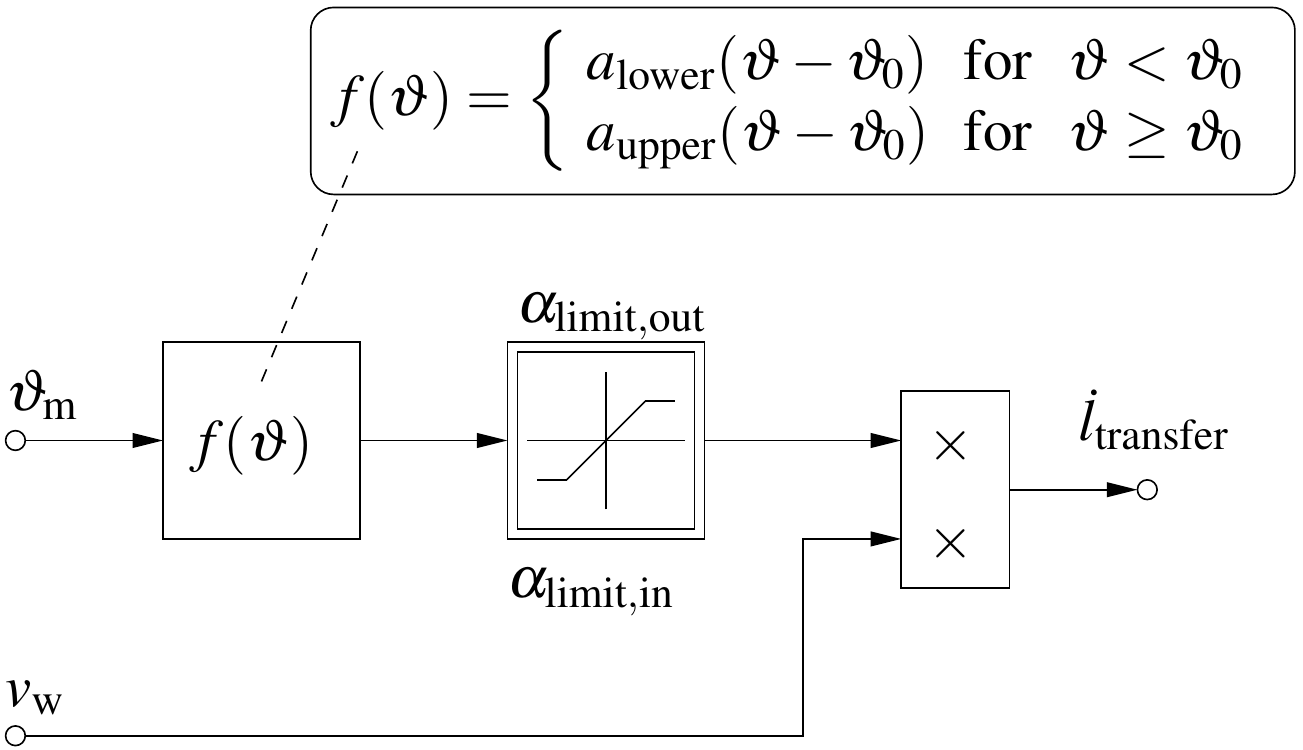}
  \caption{Figure of the winch controller representing the signal flow
  graphically.}
  \label{fig:winch_controller}
\end{figure}

The simulation, which is sketched in 
\ref{sec:cycle_optimization}, suggests a remarkable simple law for choosing the
winch speed $\dot{l}_{\rm s}$ as function of the windwindow angle $\vartheta$.
The simulated results are plotted in Fig.~\ref{fig:winch_function}.
Comparing winch speed and wind window angle, a linear dependence could
be suspected which would suggest the following ansatz for the winch controller
as given in Fig.~\ref{fig:winch_controller}.

The constant parameters are typically chosen as follows:
$\vartheta_0=1.05$\,rad, $a_{\rm lower}=-0.55$ and $a_{\rm upper}=-0.65 $.
While the $a$ values directly follow from the simulation, the $\vartheta$
value can be modified slightly for practical operation in order to
take into account line slack during the transfer phase. 
The winch speed limits are chosen as $\alpha_{\rm limit,in}=-0.5$ in order to
limit the wind window angle to approx.~$\vartheta < 1.9$\,rad
($\approx$110\,deg) during the return phase, compare
(\ref{eq:theta_winching}). The limit $\alpha_{\rm limit,out}=0.3$ is motivated by the rule of thumb given above.
For $v_{\rm w}$, either the anemometer at the ground station, or a wind
estimation algorithm output is used.

In summary, the winch controller consists of a simple function mapping the wind
window angle $\vartheta_{\rm m}$ to $v_{\rm winch}$, scaled by the wind speed
$v_{\rm w}$.
\subsection{Pattern restart winch control}
\label{sec:restart_control}
A special situation is given while flying from the static windward position
heading to target point TP3 back into the dynamical pattern. As the optimization
is based on circular orbits, which are quite different to the curve-down
maneuver from TP3 to TP1, the simulation results are not suitable for deriving
an appropriate control law.
First experimental tests have shown, that the transfer phase controller of the
previous section can be applied for the restart, albeit
the significant force peaks for lower $\vartheta$ values should be avoided.
The testing of improved controllers is subject to current research activities,
a possible restart control phase is drawn in Fig.~\ref{fig:winch_function} as
dash-dotted line.

%% file: results.tex
\section{Experimental flight results}
\label{sec:results}
\begin{figure}[h]
  \centering
  \includegraphics[width=8.9cm]{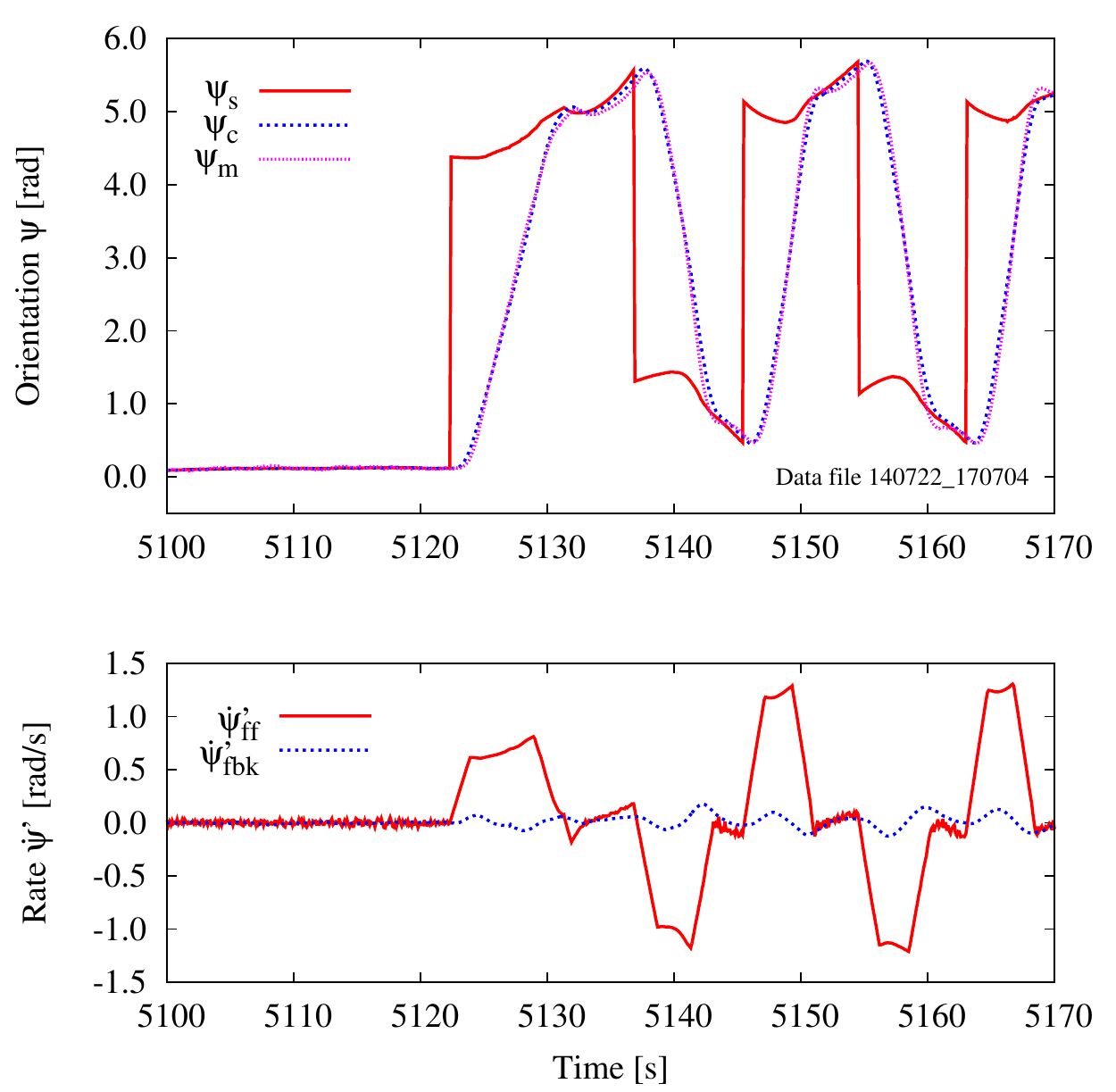}
  \caption{Principle of operation of the $\psi$ controller illustrated by
  signals of experimental flight data. The commanded $\psi_{\rm s}$ features
  steps, which are shaped by the feedforward block to get $\psi_{\rm c}$. The
  measured value $\psi_{\rm m}$ is shown for comparison.}
  \label{fig:plot_psi}
\end{figure}

\begin{figure}
  \centering
  \includegraphics[width=8.9cm]{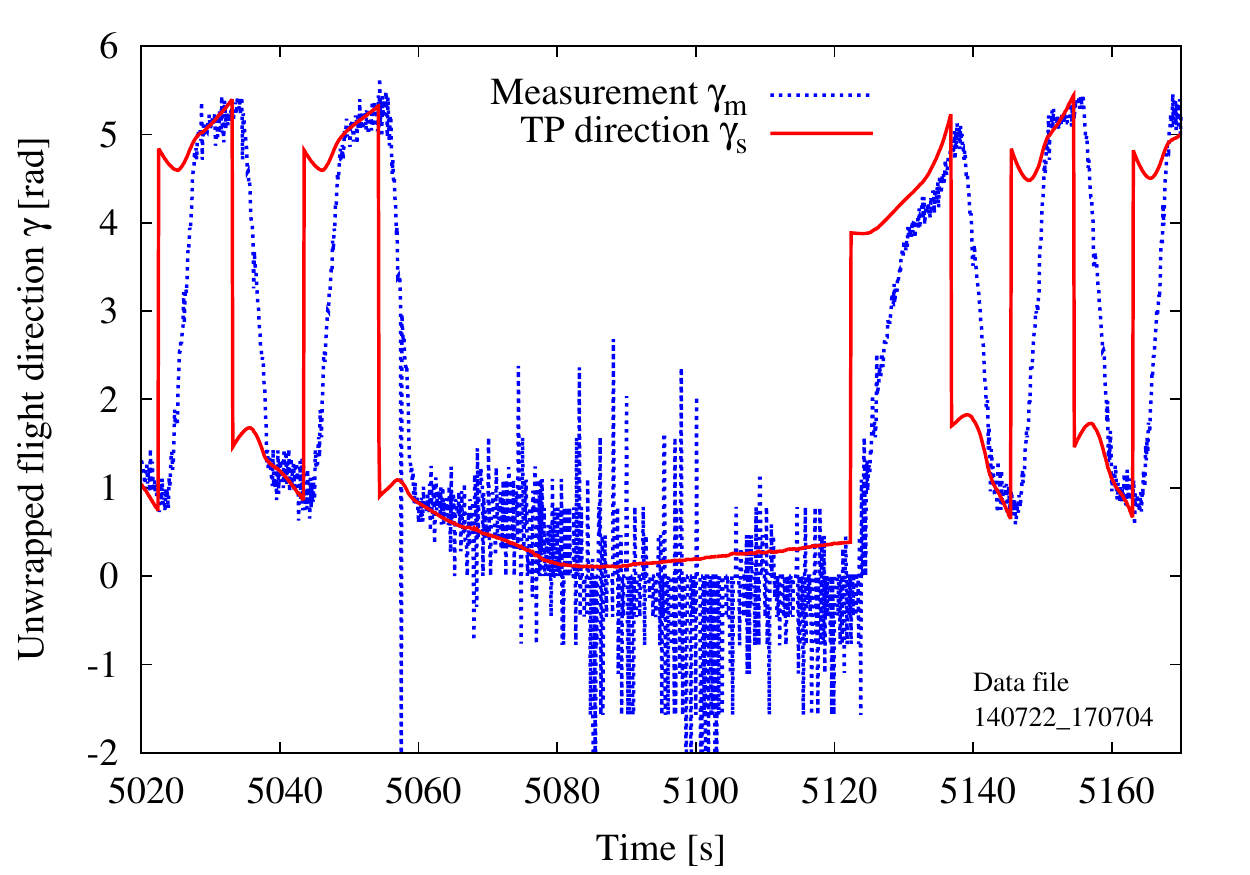}
  \caption{Measured flight directions compared to target point (TP) directions.
  The noise figure of $\gamma_{\rm m}$ originates from the difference equation
  (\ref{eq:gamma_m}). Note, that $\gamma_{\rm m}$ is plotted unfiltered for
  comparison only and not used for the control. During the return
  phase, the direction becomes more or less undefined due to the static flight.}
  \label{fig:plot_direction}
\end{figure}

\begin{figure}
  \centering
  \includegraphics[width=8.8cm]{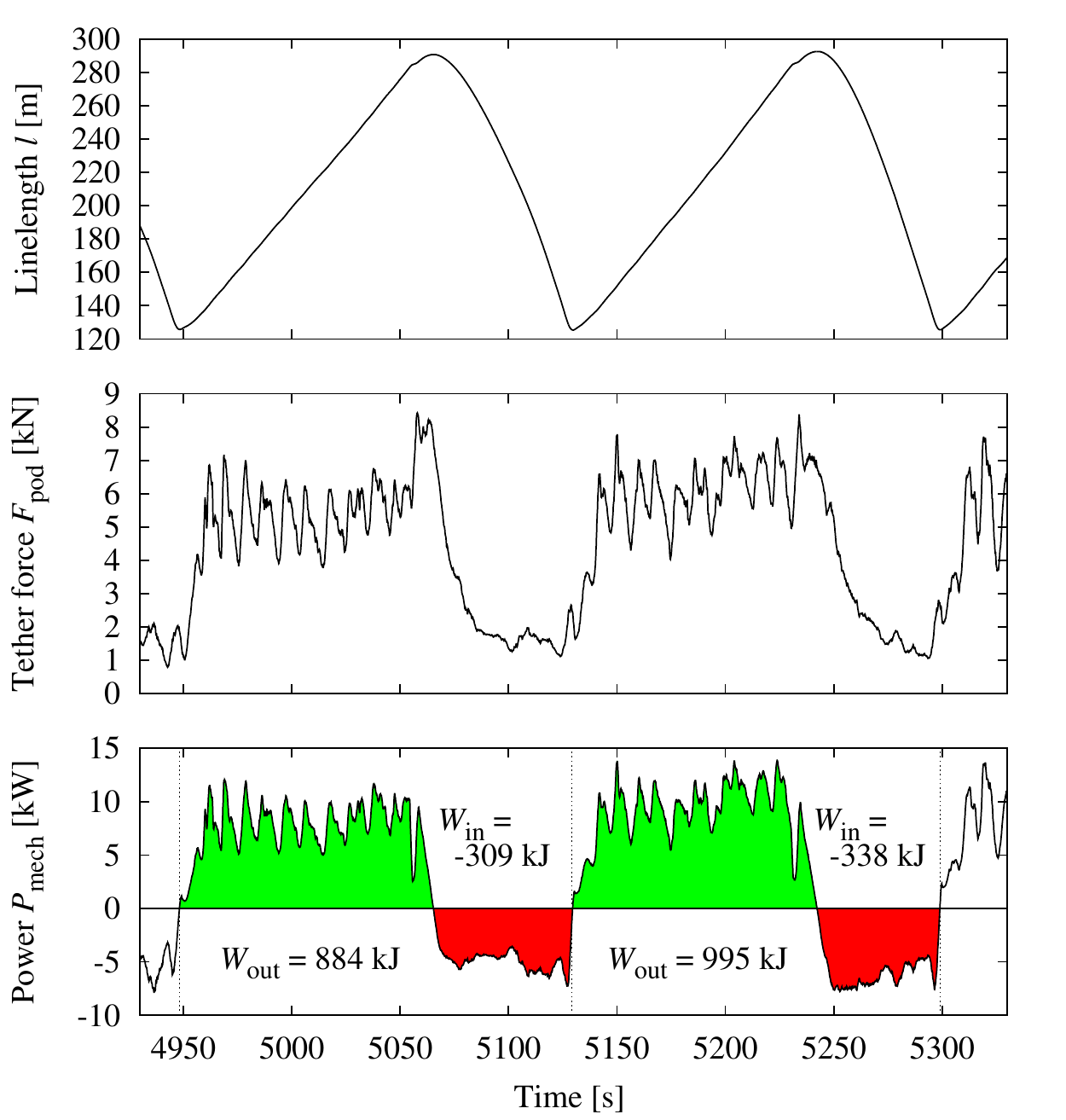}
  \caption{Performance data plot of real flight data at $v_{\rm
  w}=7.5$\,m/s (estimation for average flight altitude).
  The lower subplot shows the mechanical power computed by $P_{\rm mech}=F_{\rm
  pod}\dot{l}$. In order to estimate the electrical net power output, conversion
  efficiencies have to be taken into account. The energies $W_{\rm in}$ for
  reeling in and $W_{\rm out}$ are given seperately. The average power for the
  complete cycles is computed by 
  $\bar{P}_{\rm cycle}=(W_{\rm in}+W_{\rm out})/T$ where $T$ is the cycle time.
  The two plotted cycles yield $\bar{P}_{\rm cycle}=3.17$\,kW and
  $\bar{P}_{\rm cycle}=3.86$\,kW, respectively.
  } 
  \label{fig:power_data}
\end{figure}

In this section, experimental flight data of the SkySails small prototype, using
a 30\,m$^2$ kite at $v_{\rm w}\approx 8$\,m/s wind speed, will be discussed.
They should be regarded as first results for a proof-of-concept of the
introduced algorithms without claiming the achievement of fully optimized power
production cycles. 
\begin{enumerate}
\item Inner loop ($\dot{\psi}^\prime$-controller):
Performance results for the inner loop $\dot{\psi}^\prime$-controller are given
in Fig.~\ref{fig:plot_psidot}.
\begin{figure}
  \centering
  \includegraphics[width=8.9cm]{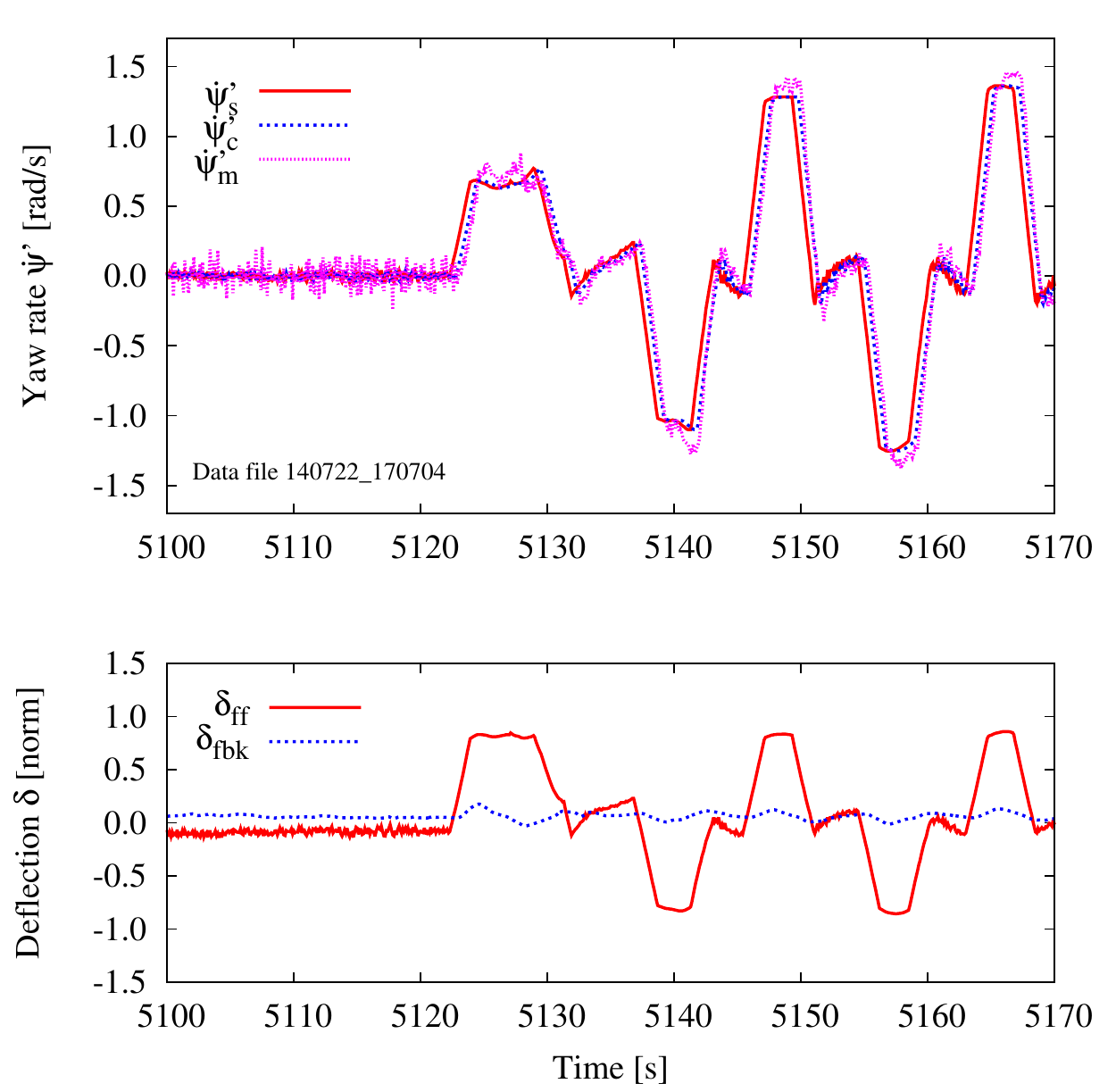}
  \caption{Principle of operation of the $\dot{\psi}^\prime$ controller
  illustrated by signals of experimental flight data.
  Upper plot: comparison of set value $\dot{\psi}_{\rm s}^\prime$ (solid),
  reference value $\dot{\psi}_{\rm c}^\prime$ (dashed) and measured value
  $\dot{\psi}_{\rm m}^\prime$ (dotted).
  Lower plot: feedforward (solid) and feedback (dashed) outputs of the
  $\dot{\psi}^\prime$ controller.}
  \label{fig:plot_psidot}
\end{figure}
In order to review the controller performance, 
the set ($\dot{\psi}_{\rm s}^\prime$) 
and reference $\dot{\psi}_{\rm c}^\prime$ values for the turn rate are compared
to the measured turn rates $\dot{\psi}_{\rm m}^\prime$.
The excellent agreement of the measured turn rate shows, that system
constraints are met.
Note, that the ratio of feedback to feedforward commands is quite small, which
indicates, that the feedback is only necessary for correcting small
perturbations and uncertainties.
This further validates the turn-rate law (\ref{eq:ctl_psidot_trl}) and
(\ref{eq:extended_trl}), on which the feedforward path is based.
\item Outer loop ($\psi$-controller): The respective controller signals
are shown in Fig.~\ref{fig:plot_psi}.
Note, that although switching of target points introduces steps to the commanded
$\psi_{\rm s}$, the output quantity $\psi_{\rm c}$ is properly shaped subject to
limited steering deflection $\delta_{\rm s}$ and speed $\dot{\delta}_{\rm p}$.
An accurate analogy of the estimated orientation $\psi_{\rm m}$ to the reference
value $\psi_{\rm c}$ can be observed, resulting in the fact, that dynamics
is mainly controlled by the feedforward part. 
Hence, a proper design of the inner loop as well as validity of the dynamics
given in Sect.~\ref{sec:plant_crossterm} can be stated.
\item Guidance: In order to evaluate the flight direction control, the direction
to the active target point is compared to the measured flight direction as plotted
in Fig.~\ref{fig:plot_direction}.
The measured directions are computed by differences $\Delta\varphi$,
  $\Delta\vartheta$ between consequent samples using (\ref{eq:def_gamma}), i.e.
  \begin{equation}
    \gamma_{\rm m}=\arctan(-\Delta\varphi
    \cos\vartheta,\Delta\vartheta)\label{eq:gamma_m}
  \end{equation}
For the return phase with insignificant kinematic motion, $\gamma_{\rm m}$
becomes more or less undefined and unsuitable as direct control input. 
In order to handle this issue, a regularization, as proposed in
\cite{Zgraggen2013b,Zgraggen2014b}, could be used. 
In contrast, we decided to base the direction control on $\psi$ as introduced in
Sect.~\ref{sec:flight_direction}.
The excellent agreement of directions during the dynamic flight apart from the
continuous $\gamma_{\rm m}$ following the step in $\gamma_{\rm s}$ during
curves, demonstrates, that the flight direction control works satisfactorily and
relation  flights characterized by a step in $\gamma_{\rm s}$ and
relation (\ref{eq:flight_direction}) between $\gamma$ and $\psi$ holds.
\end{enumerate}
Finally, this section will be closed with a discussion
of experimental power generation data as shown in Fig.~\ref{fig:power_data}.
As reference value for the generated power, the Loyd's limit is used, which
corresponds to a continuous crosswind flight and optimal winch speed for that
situation. This limit is calculated 
assuming, that the airpath speed in crosswind condition is determined by the
back wind reduced by winching and the glide ratio as $v_{\rm a}=E (v_{\rm
w}-\dot{l})$. Using (\ref{eq:P_bar}) and determining the maximum by varying
$\dot{l}$, one finds the above mentioned thumb rule $\dot{l}=v_{\rm w}/3$
and for the power:
\begin{equation}
  P_{\rm Loyd} = \frac{\varrho C_{\rm R} A}{2} \frac{4E^2}{27}v_{\rm w}^3
  \label{eq:loyds_limit}
\end{equation}
For the flight test conditions $A=0.7\cdot 30$\,m$^2$,
$\varrho=1.2$\,kg/m$^3$, $C_{\rm R}=1.0$, $E=5.0$ and $v_{\rm w}=7.5$\,m/s
(estimated at mean flight altitude), $P_{\rm Loyd}\approx 19.7$\,kW
is obtained.
Evaluating a typical cycle, as shown in Fig.~\ref{fig:power_data}, yields an
average power of $\bar{P}=3.5$\,kW, hence $\bar{P}=0.18 P_{\rm Loyd}$,
which is about 75\% of the expected value due to simulation results given in
\ref{sec:cycle_optimization}. 

Taking into account, that the simulation does not
consider the significant losses due to curve flights and real conditions coming
along with gusts, the theoretical simulation based on the simple model and the
experimental finding can be regarded as consistent.
Hence, although neither precision measurements nor an extended optimization of
the power output were in the scope of the presented prototype setup, the
presented model as well as the controller setup could be regarded as solid bases
for further development steps. Especially the trajectory of recycling back into
the power phase, but also the whole power phase are candidates for significant
improvements in the near future.

%% file: summary.tex
\section{Summary and future work}
\label{sec:summary}
We have presented a complete control setup for autonomous power generation
flying tethered kites at constant glide ratio in the pumping cycle scheme.
The control structure design is based on and motivated by a simple model
for the system dynamics.
The discussed flight data with a small-scale demonstrator illustrates, that
the implemented target point control allows to fly reliably the pattern
eight-down, which is an ambitious scheme, but allows for significantly
higher power generation efficiency.

The winch control has been implemented as simple state-feedback and tailored
for an efficient transfer phase. Although it is quite rudimentary, it can be
utilized during all flight phases with energy generation results not fully optimized
but already remarkable.

With respect to future work, there are two major fields for further
developments. First, a long-term autonomous operation requires a high
level of robustness of the control system under extreme environmental
conditions, which basically involve extreme gusts and temporary untethered states.
Especially, the development of estimation and filtering algorithms of sensor
values is a research field on its own, which has been kept out of this paper in
order to focus on the general scheme and control. Extended results will be
published elsewhere.
Second, it should be noted, that the whole field of
optimizing the power output has only be scratched by the surface in this paper.
An extended understanding of optimization criteria, which also take into
account different wind conditions at different altitudes, has to be established. A
further major challenge will arise from the implementation of robust
operational algorithms smartly adapting to varying environmental conditions.
Currently, only few theoretical proposals \cite{diwale2014optimization} and
experimental results on that issue have been reported, see e.g.~the maximum power point
searching stepping algorithm \cite{Zgraggen2014a}, which adapts to varying wind
directions.

The final goal
of implementing fully autonomous airborne wind energy power plants additionally
demands for control systems for starting and
landing of the kite.
The differences in system dynamics at short tether lengths 
lead to different control approaches, which are subject to current research
activities.

%% file: appendix.tex
\section{Optimization of power generation cycles}
\label{sec:cycle_optimization}
In order to get a principal idea of how to perform transfer and return phases
w.r.t.~winch speed efficiently with the setup of {\em constant} glide ratio,
 a heuristically motivated optimization has been applied to the simple model,
 which will be briefly presented in the following.

\begin{figure}[h]
  \centering
  \includegraphics[width=8.8cm]{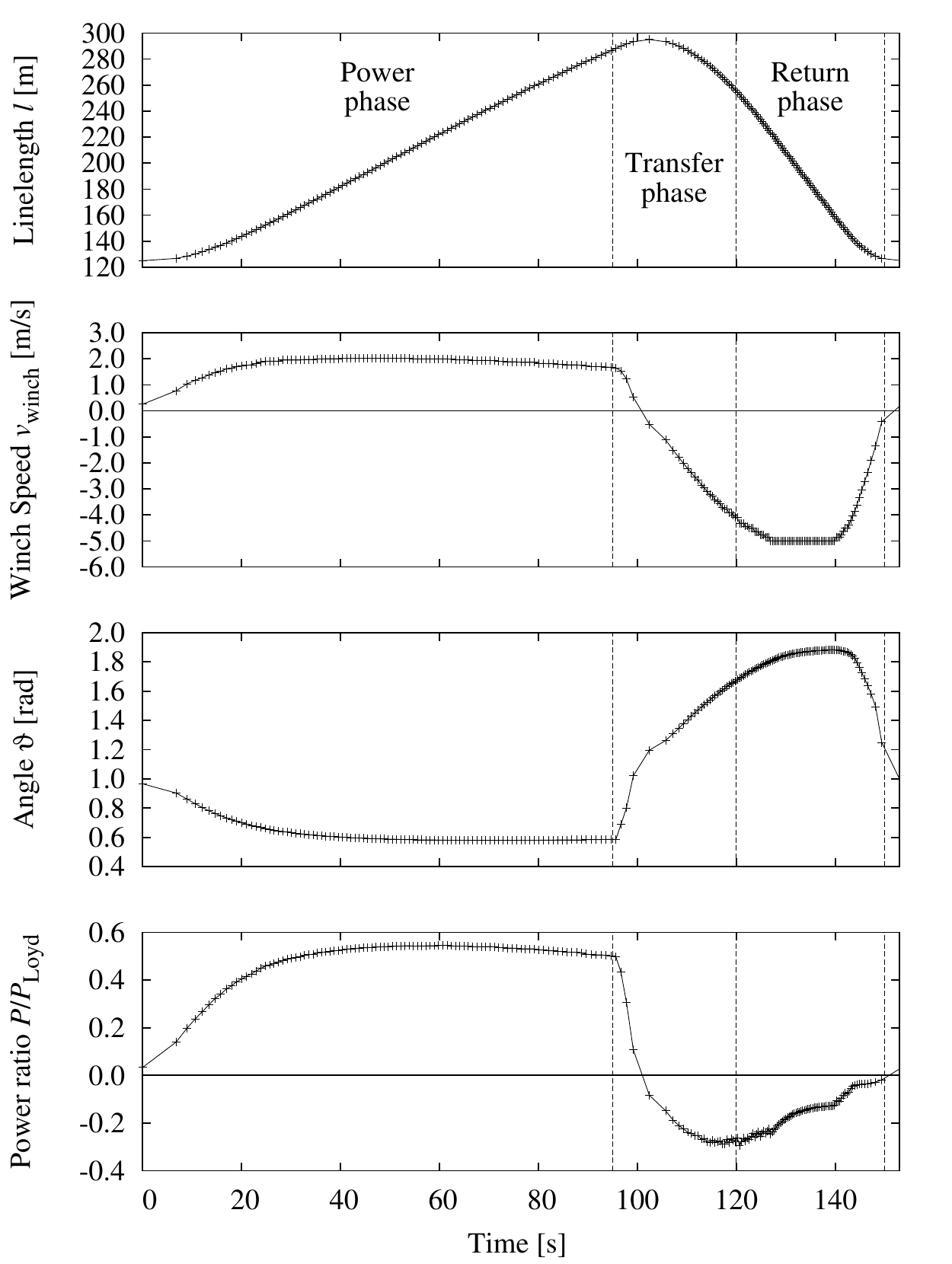}
  \caption{Optimization results for $v_{\rm w}=10$\,m/s. The plots comprise one
  cycle and single phases are separated by vertical dashed lines.
  The power (lower subplot) is normalized to the Loyd's limit $P_{\rm
  Loyd}$, compare (\ref{eq:loyds_limit}), and the average value is given by
  $\bar{P}=0.24 P_{\rm Loyd}$.}
  \label{fig:plot_simulation}
\end{figure}

The quantity to optimize is the average power of a complete cycle of duration $T$
\begin{equation}
  \bar{P} = \frac{1}{T}\int\limits_0^T dt\,\dot{l}(t) F(t) 
\end{equation}
using the expression for the tether force
\begin{equation}
  F(t) = \frac{\varrho}{2} C_{\rm R} A v_{\rm a}^2(t)
\end{equation}
with $\varrho$ the air density, $A$ the projected kite area and $C_{\rm R}$ the
force coefficient, one obtains for the average power
\begin{equation}
  \bar{P} =  
  \frac{\varrho C_{\rm R} A}{2T}\int\limits_0^T dt\, \dot{l}(t) v_{\rm a}^2(t)
  \label{eq:P_bar}
\end{equation}
As it is reasonable to vary $\psi(t)$ and $\dot{l}(t)$, these are considered as
{\em input functions}. As consequence, a functional
$\bar{P}: \{\psi(t),\dot{l}(t),T\} \mapsto \bar{P}
[\psi(t),\dot{l}(t),T]$ can be defined by solving the equations of motion (\ref{eq:eqm_theta}),
(\ref{eq:eqm_phi}), (\ref{eq:eqm_va}) and using (\ref{eq:P_bar}).
The optimization problem can now be stated:
\begin{equation}
  {\rm maximize} \,\, \bar{P} [\psi(t),l(t),T] 
\end{equation}
by variation of $\{\psi(t), l(t), T\}$ subject to the following
constraints:
\begin{itemize}
  \item Periodic boundaries: $l(T)=l(0)$ and $\psi(T)=\psi(0)$. Further, the
  initial condition for (\ref{eq:eqm_theta}) must also be periodic
  $\vartheta(0)=\vartheta(T)$.
  \item There is only one reel-in and one reel-out phase, i.e.~$\dot{l}(t)$ has
  only two roots for $0\leq t<T$. In addition, the operational range is given,
  $l_{\rm min}\leq l \leq l_{\rm max}$ and these range limits must be reached.
  \item $\alpha_{\rm limit,in} \leq (\dot{l}/v_{\rm w}) \leq \alpha_{\rm
  limit,out}$
  \item $0 \leq \psi(t) \leq \psi_{\max}$, $\psi_{\rm max}$ and the $\varphi$
  motion is free and not subject to periodic boundary condition through cycles.
  These condition lead to circular orbits (also below the surface) instead
  of figure-eights during the power phase. The $\psi_{\rm max}$ value can
  directly by regarded as force control as it determines the equilibrium wind window position as discussed in
  detail in \cite{Erhard2012a}. This rough approximation of figure eights
  significantly simplifies the optimization algorithm and is adequate for a
  first approach. Future work should include proper figure eights and
  consider losses due to curve flights \cite{Costello2013a}.
\end{itemize}
It should be stated, that the
simulation has been performed for $v_{\rm w}=10$\,m/s. Winch speeds for other
wind speeds can be easily obtained by scaling the results with $v_{\rm w}$
accordingly.
Simulation curves are shown in Fig.~\ref{fig:plot_simulation}.
As a result for the transfer and return phase, a similarity of winch speed
$v_{\rm winch}$ and angle $\vartheta$ can be recognized, which motivates the
implementation of a winch controller as linear functions $v_{\rm winch}=v_{\rm
winch}(\vartheta)$, compare Fig.~\ref{fig:winch_function}.
Note, that in the first part of the transfer phase the optimization
yields $v_{\rm winch}>0$ and thus suggests utilization of the high tether
forces for energy production. Winching in starts for elevation angles higher
than $\vartheta_0 \approx 1.05$\,rad.

Finally, it has to be remarked, that the results in the
appendix have to be considered as a first step in tackling the cycle
optimization issue.
They should be regarded as illustration of the basic idea rather than a
complete treatment of the problem. Hence, further extended
investigation of this optimization problem is recommended, which may lead to
major modifications of the winch control strategy presented in this paper.